\documentclass[a4paper,english,12pt,preprint]{iopart}
\usepackage[T1]{fontenc}
\usepackage[latin1]{inputenc}
\usepackage{graphicx}
\usepackage{amssymb}

\makeatletter

\usepackage[]{times}
\usepackage{ae}
\usepackage{cite}

\usepackage{babel}
\makeatother
\begin{document}

\title{Spectral and polarization effects in deterministically nonperiodic
multilayers containing optically anisotropic and gyrotropic materials}

\author{S. V. Zhukovsky$^{1}$, V. M. Galynsky$^{2}$}

\address{$^{1}$ Physikalisches Institut, Universit\"at Bonn, Nussallee 12,
53115 Bonn, Germany}

\address{$^{2}$ Physics Dept., State University of Belarus, Fr.~Skaryna
Ave. 4, 220080 Minsk, Belarus}

\ead{sergei@th.physik.uni-bonn.de}

\begin{abstract}
Influence of material anisotropy and gyrotropy on optical properties
of fractal multilayer nanostructures is theoretically investigated.
Gyrotropy is found to uniformly rotate the output polarization for
bi-isotropic multilayers of arbitrary geometrical structure without
any changes in transmission spectra. When introduced in a polarization
splitter based on a birefringent fractal multilayer, isotropic gyrotropy
is found to resonantly alter output polarizations without shifting
of transmission peak frequencies. The design of frequency-selective
absorptionless polarizers for polarization-sensitive integrated optics
is outlined.

\end{abstract}


\submitto{\JOA}

\maketitle

\section{Introduction\label{sec:Introduction}}

Periodic nanostructured media, also known as photonic crystals (see,
e.g., \cite{Joannopoulos,Sakoda} and references therein), have put
forth a wide range of possible applications in optical communication,
optoelectronics and optical means of data transmission. In addition,
photonic crystals have inspired a lot of fundamental research on light-matter
interaction, since electromagnetic wave propagation phenomena in strongly
inhomogeneous media have been for the first time available for direct
theoretical investigation.

Moreover, during the recent decade it was shown that further alteration
of topology of inhomogeneous media leads to an even wider choice of
optical materials. Within this scope, \emph{non-periodic deterministic
(NPD) structures} have been extensively studied during the recent
decade. Namely, quasiperiodic \cite{Kohmoto,Macia,Alb} and fractal
\cite{PRE,SPIE-icono,EPL} nanostructures are known to possess distinct
optical properties, not present in either periodic or disordered systems.
The most interesting among these are spectral self-similarity and
critical eigenstates in Fibonacci quasiperiodic lattices \cite{Kohmoto}
and spectral scalability in Cantor fractal multilayers \cite{PRE},
which was shown to directly result from geometrical self-similarity
\cite{EPL}. 

However, nearly in all the research in this field only optically \emph{isotropic}
constituent materials are considered. On the other hand, there have
been recent reports with investigation of \emph{anisotropic} materials
embedded into a \emph{periodic} structure. For example, it was shown
that anisotropy can cause even a one-dimensional (1D) multilayer structure
to exhibit negative refraction \cite{Liu,Antonio}, a phenomenon earlier
attributed exclusively to higher-dimensional systems, as well as omnidirectional
total transmission \cite{Liu,AntonioJOSA}. In addition, interesting
results have been recently obtained with weakly modulated periodic
structures made of anisotropic and gyrotropic materials \cite{Makarov04}.
Extensive ongoing analytical research of anisotropic and gyrotropic
multilayers \cite{BorzdBiAn97,Borzdov99,BorzdBiAn00}, as well as
studies of a simpler case of chiral media in 2D \cite{Chiral2D} and
3D \cite{Chiral3D} photonic crystals, is also restricted to periodic
systems.

From what has been said, there is a need to ``bridge the gap'' and
combine anisotropy and non-periodicity in the same structure. It is
of particular interest to find out how optical anisotropy of constituent
materials can affect the spectral properties caused by deterministic
non-periodic geometry.

\begin{figure}[b]
\begin{center}\includegraphics[%
  bb=0bp 30bp 353bp 279bp,
  scale=0.3]{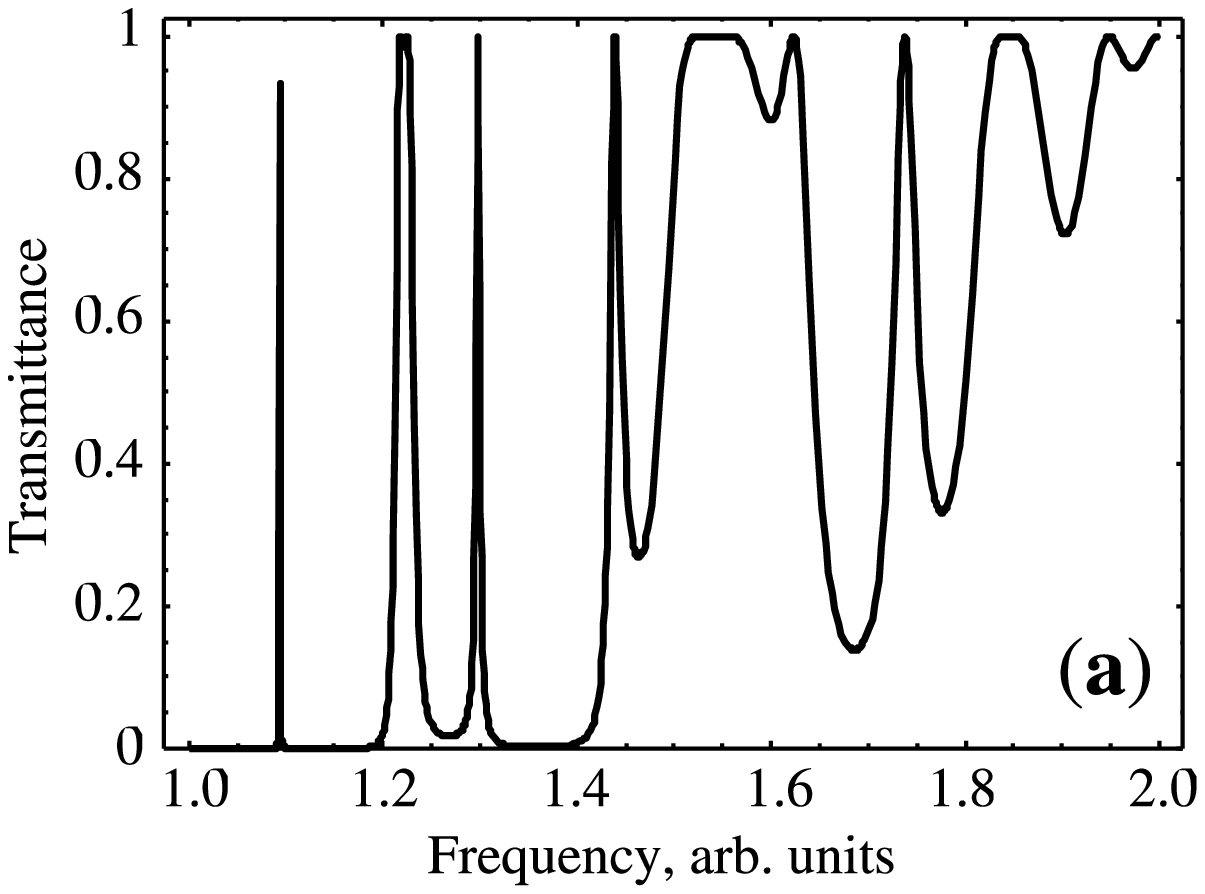}\includegraphics[%
  bb=0bp 30bp 446bp 279bp,
  scale=0.3]{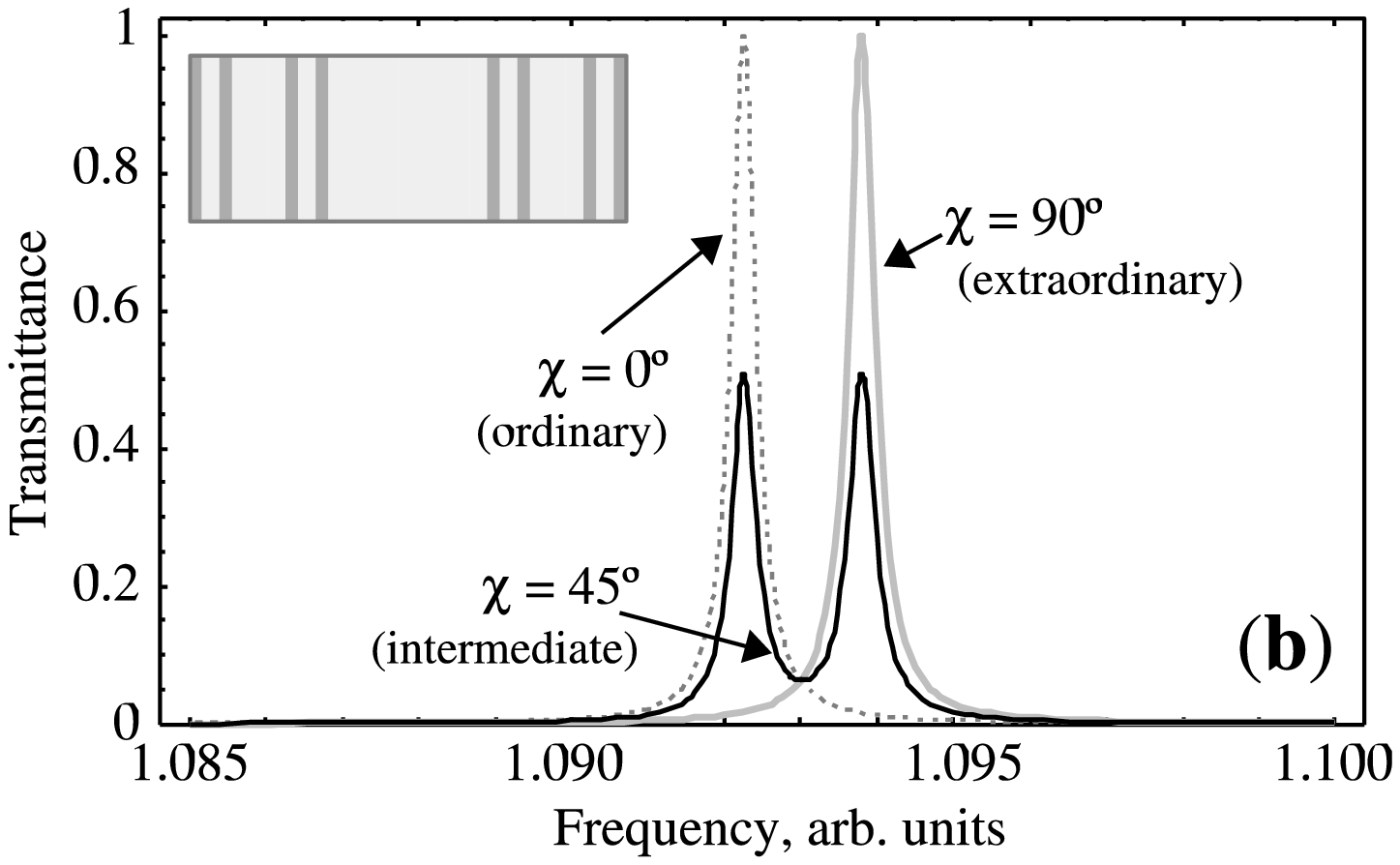}\includegraphics[%
  bb=0bp 30bp 446bp 279bp,
  scale=0.3]{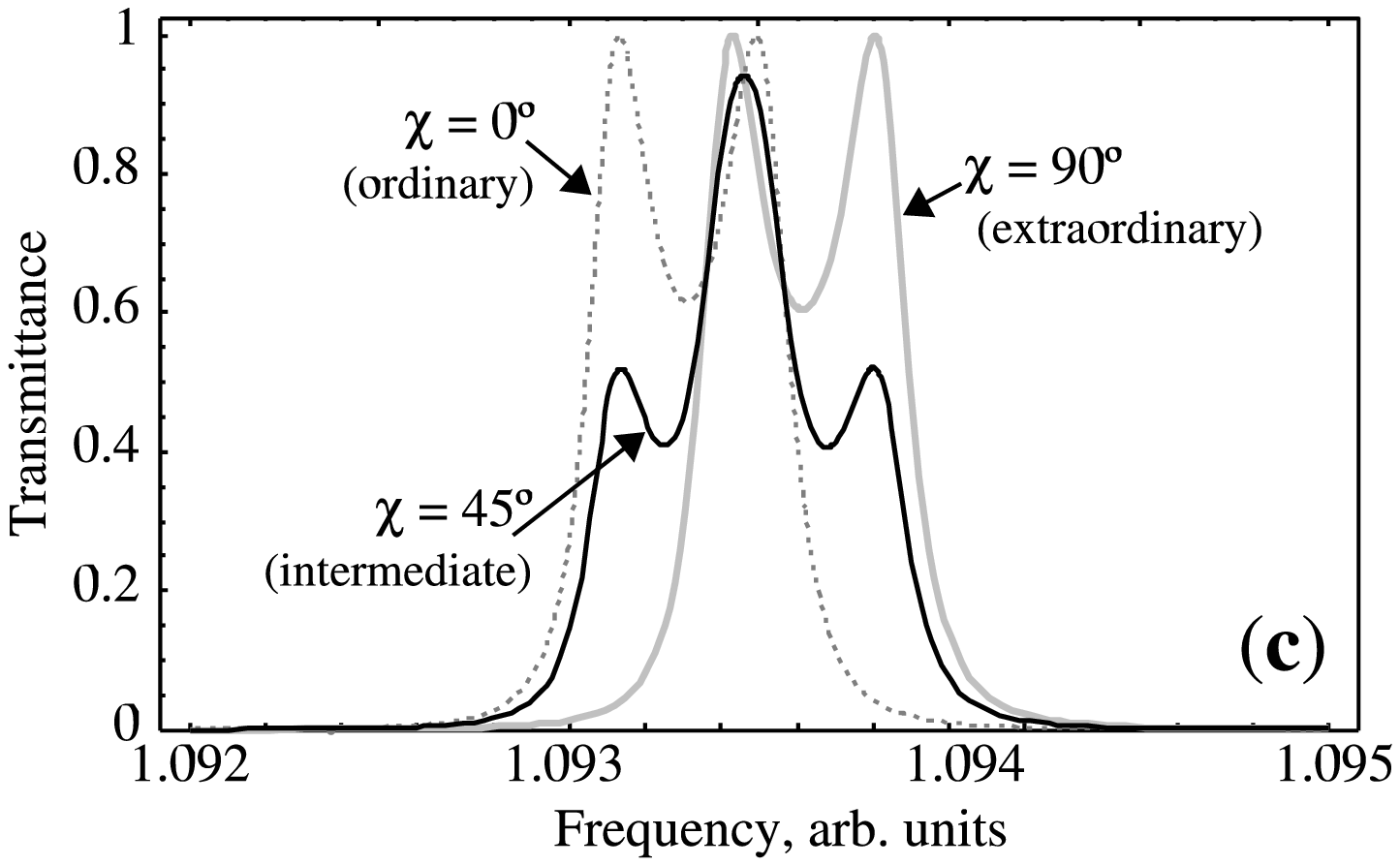}\end{center}

\caption{\label{fig:splitting}A portion of spectrum of a 27-layer Cantor
fractal multilayer structure as a typical spectrum of an NPD structure
\textbf{(a)} along with the image of one peak \textbf{(b)} at different
orientations of input wave polarization. The peak exhibits polarization-induced
splitting \cite{Warsaw}. Sometimes, a doublet may split in such a
way that matching occurs \textbf{(c)}. The inset shows the structure
geometry, which can be written as $gagaaagagaaaaaaaaagagaaagag$,
$g$~being GaN~($\varepsilon_{o}=5.685$,~$\varepsilon_{e}=5.892$)
and $a$~being air quarter-wave layers. Frequency is normalized to~$\omega_{0}$
defined as $n_{g}d_{g}=n_{a}d_{a}\equiv\lambda_{0}/4=\pi c/2\omega_{0}$
\cite{Warsaw}.}
\end{figure}

Our earlier communication \cite{Warsaw} showed that for any multilayer
structure exhibiting sharp resonance peaks inside a band gap (generally,
a characteristic feature of NPD geometry) anisotropy results in polarization-induced
peak splitting (figure~\ref{fig:splitting}a,b). In this regime,
the structure acts as a nanosized absorptionless frequency-selective
polarizer, which can in itself find certain applications. Moreover,
in the case when a doublet undergoes polarization-induced splitting
in such a way that the components are matched (figure~\ref{fig:splitting}c),
the structure acts as a quarter-wave retarder (QWR) and converts a
linearly-polarized beam into circularly-polarized and vice versa.
The size of such a retarder is smaller than that of a bulk QWR, and
the content of birefringent material is more than 10 times decreased,
which suggests enhancement of effective birefringence when anisotropic
material in organized in deterministic non-periodic manner. 

Of the geometries in question, fractal structures are of special interest
because their spectra can provide resonance peaks with controlled
multiplicity due to sequential peak splitting \cite{PRE}.

In this paper, we move on to further analyze the optical properties
of non-periodic media containing optically anisotropic constituent
materials. Along with conventional birefringence, we analyze the effects
caused by optical activity (or \emph{gyrotropy}) \cite{Fedorov}.
We show numerically as well as analytically that in the case of bi-isotropic
multilayers (isotropic optical activity introduced into an initially
isotropic medium) gyrotropy does not cause polarization-induced splitting
-- contrary to what might have been expected. In fact, gyrotropy even
does not manifest itself in the spectral characteristics of the multilayer,
its sole influence being a geometry-independent polarization plane
rotation of the transmitted wave. Furthermore, we show that optical
activity can be made use of to modify the polarization sensitivity
and the output polarization of a birefringent polarizer described
above, meaning both polarization orientation and ellipticity. Possible
applications for integrated optics and optical engineering are also
outlined.

To account for gyrotropy, we make use of coordinate-free operator
formalism for the general case of bi-anisotropic multilayers \cite{FedorovNew,BBL}.
It is outlined briefly in Section~\ref{sec:Theoretical}. The simplest
case of a multilayer with bi-isotropic layers is analytically and
numerically investigated in Section~\ref{sec:Biisotropic}. It is
shown that gyrotropy can only provide polarization plane rotation
completely independent of structure geometry. The combination of isotropic
optical activity and uniaxial birefringence is analyzed in Section~\ref{sec:Gyrotropy},
and it is shown that in this case gyrotropy can be used to control
the output polarization of an integrated optical polarizer. Section~\ref{sec:Discussion}
summarizes the paper.

\section{Evolution operator for anisotropic and gyrotropic media\label{sec:Theoretical}}

Before proceeding, it is necessary to briefly touch upon the theoretical
approaches used. Contrary to the case with isotropic media when common
transfer matrix methods with scalar field values would suffice, anisotropic
and gyrotropic media cannot be regarded without accounting for polarization
effects. Such effects include, for instance, polarization coupling
at layer interfaces, so in the general case wave polarization cannot
be separated into independent states and full vector calculations
have to be performed.

Among the methods for such calculation, we have used the coordinate-free
operator method (also known as Fedorov's covariant approach \cite{Fedorov,FedorovNew,FedorovEng}),
developed for stratified systems in \cite{BBL}. In this method most
equations remain in compact form containing only the inherent parameters
of the structure. This provides flexibility and facilitates analytical
investigation of the equations used.

To account for arbitrary anisotropy and gyrotropy, the following form
of material equations is used \cite{BBL}:

\begin{equation}
\mathbf{D}=\varepsilon\mathbf{E}+i\alpha\mathbf{H},\quad\mathbf{B}=\mu\mathbf{H}+i\beta\mathbf{E}.\label{eq:mat}\end{equation}

Here $\varepsilon$ and $\mu$ are dielectric permittivity and magnetic
permeability, respectively. In optically anisotropic media, $\varepsilon$
is a tensor rather than just a scalar coefficient. In materials possessing
magnetic anisotropy, $\mu$ can also be a tensor. The quantities $\alpha$~and~$\beta$,
also tensorial in the general case, are called \emph{gyration pseudotensors}
and are responsible for the material's optical activity. In non-absorbing
media they must satisfy the relation $\beta=-\alpha^{\dagger}$ (Hermitian
conjugation). The material equations~(\ref{eq:mat}) can in theory
encompass all possible cases of anisotropy and gyrotropy \cite{Fedorov}.

\begin{figure}
\begin{center}\includegraphics[%
  scale=0.3]{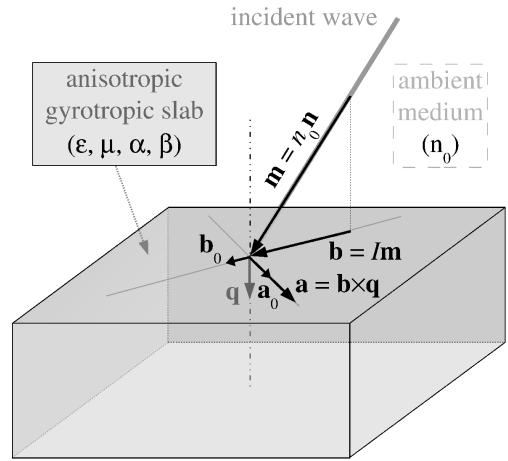}\end{center}

\caption{\label{fig:basis}Illustration of an ambient medium / layer interface:
on the definition of a natural basis in~(\ref{eq:basis}).}
\end{figure}

Considering a plane wave incident from isotropic ambient medium with
refractive index $n_{0}$ onto a slab of a material with parameters
given in (\ref{eq:mat}), one can define the wave by its vector of
refraction $\mathbf{m}=n_{0}\mathbf{n}$, $\mathbf{n}$~being the
vector of wave normal (figure~\ref{fig:basis}). Analogously, the
boundary can be characterized by the unit normal vector~$\mathbf{q}$
and the corresponding projection operator

\[
I=\mathbf{1}-\mathbf{q}\otimes\mathbf{q}=-(\mathbf{q}^{\times})^{2},\]
which projects any vector into the interface plane. Here the notation
$\mathbf{u}\otimes\mathbf{v}$ denotes an outer product (also termed
the \emph{dyad} \cite{BBL}) between two vectors $\mathbf{u}$~and~$\mathbf{v}$,
which is a tensor with elements in the form $(\mathbf{u}\otimes\mathbf{v})_{ij}=u_{i}v_{j}$,
and the symbol~$\mathbf{q}^{\times}$ stands for an antisymmetric
tensor defined via vector cross product as $(\mathbf{q}^{\times})\mathbf{u}=\mathbf{q}\times\mathbf{u}$,
or via fully antisymmetric Levi-Civita's pseudotensor~$\mathcal{E}_{ijk}$
as $\left(\mathbf{q}^{\times}\right)_{jk}=\mathcal{E}_{ijk}\mathbf{q}_{i}$.

Using these elements as a reference, one can further introduce the
vectors\begin{equation}
\mathbf{b}=I\mathbf{m},\quad\mathbf{a}=\mathbf{b}^{\times}\mathbf{q}.\label{eq:basis}\end{equation}

The three vectors $(\mathbf{a}_{0}\equiv\mathbf{a}/|\mathbf{a}|,\mathbf{b}_{0}\equiv\mathbf{b/}|\mathbf{b}|,\mathbf{q})$
are of unit length and mutually orthogonal. Thus they form a basis
in space (see figure~\ref{fig:basis}). Note that at normal incidence
$\mathbf{a}=\mathbf{b}=0$, which results from axial symmetry of such
a system. In this particular case only one of the basis vectors can
be introduced naturally, while $\mathbf{a}_{0}$~and~$\mathbf{b}_{0}$
are subject to arbitrary choosing.

It was shown earlier \cite{BBL} that Maxwell's equations with~(\ref{eq:mat})
can be transformed into an operator differential equation

\begin{equation}
\begin{array}{l}
\frac{d}{dx}\mathbb{W}(x)=ik\mathbb{MW}(x),\\
\\\mathbb{W}(x)=\left[\begin{array}{c}
I\mathbf{H}(x)\\
\mathbf{q}^{\times}\mathbf{E}(x)\end{array}\right],\quad\mathbb{M}=\left[\begin{array}{cc}
A & B\\
C & D\end{array}\right].\end{array}\label{eq:diff}\end{equation}
 Here~$\mathbb{M}$ is a characteristic block matrix (or characteristic
operator) whose elements are operators dependent on the material parameters
of the slab. It can be represented by a~$6\times6$ matrix. However,
the block matrix notation allows to discriminate between real space
{[}where we have introduced the vectors~(\ref{eq:basis}){]} and
``artificial'' $2\times2$~space related to block vectors~$\mathbb{W}$,
introduced because electric and magnetic field evolution has to be
accounted for simultaneously. To differentiate from real-space vectors
and operators, block vectors and matrices are denoted by light capital
symbols.

So the components of the characteristic matrix in~(\ref{eq:diff})
are operators acting in real space. For their full form one can refer
to \cite{BBL}. Throughout this paper, normal incidence will be considered
both in analytical and in numerical investigations. This makes certain
sense from the experimental point of view because otherwise gyrotropy
is known to be completely shadowed by common birefringence. In this
case, 

\begin{equation}
\begin{array}{l}
A=i\mathbf{q}^{\times}\alpha I+[\mathbf{q}^{\times}\varepsilon\mathbf{q}\otimes\mathbf{v}_{3}+i\mathbf{q}^{\times}\alpha\mathbf{q}\otimes\mathbf{v}_{1}],\\
B=-\mathbf{q}^{\times}\varepsilon\mathbf{q}^{\times}+[\mathbf{q}^{\times}\varepsilon\mathbf{q}\otimes(\mathbf{q}^{\times}\mathbf{v}_{4})+i\mathbf{q}^{\times}\alpha\mathbf{q}\otimes(\mathbf{q}^{\times}\mathbf{v}_{2})],\\
C=I\mu I+[iI\beta\mathbf{q}\otimes\mathbf{v}_{3}+I\mu\mathbf{q}\otimes\mathbf{v}_{1}],\\
D=-iI\beta\mathbf{q}^{\times}+[iI\beta\mathbf{q}\otimes(\mathbf{q}^{\times}\mathbf{v}_{4})+I\mu\mathbf{q}\otimes(\mathbf{q}^{\times}\mathbf{v}_{2})],\end{array}\label{eq:abcd_n}\end{equation}

Here the four auxiliary vectors $\mathbf{v}_{k}$ are introduced as
follows.\begin{equation}
\begin{array}{ll}
\mathbf{v}_{1}=-Q\mathbf{q}(\varepsilon_{q}\mu+\beta_{q}\alpha)I, & \quad\mathbf{v}_{2}=iQ\mathbf{q}(\beta_{q}\varepsilon-\varepsilon_{q}\beta)I,\\
\mathbf{v}_{3}=iQ\mathbf{q}(\alpha_{q}\mu-\mu_{q}\alpha)I, & \quad\mathbf{v}_{4}=-Q\mathbf{q}(\mu_{q}\varepsilon+\alpha_{q}\beta)I.\end{array}\label{eq:v1234_n}\end{equation}

In these expressions, the subscript~$q$ means multiplication by~$\mathbf{q}$
on both sides ($\varepsilon_{q}=\mathbf{q}\varepsilon\mathbf{q}$,
etc.), $Q=(\varepsilon_{q}\mu_{q}+\alpha_{q}\beta_{q})^{-1}$~being
the normalizing factor.

Having established the characteristic operator (\ref{eq:abcd_n})
for a layer with thickness~$d$, one can proceed to solve the equation~(\ref{eq:diff})
by taking the matrix exponential\begin{equation}
\mathbb{W}(0)=\mathbb{PW}(d),\quad\mathbb{P}=\exp\left[i\frac{\omega}{c}d\mathbb{M}\right].\label{eq:prop}\end{equation}

Here the operator~$\mathbb{P}$ is termed the \emph{propagator} or
evolution operator. It completely determines the wave propagation
in one layer made of any material. Once known for each layer of an
$N$-layer structure, the propagator for the whole system can be obtained
through operator multiplication:\begin{equation}
\mathbb{P}=\mathbb{P}_{N}\mathbb{P}_{N-1}\cdots\mathbb{P}_{1}.\label{eq:multipl}\end{equation}

The resulting evolution operator~$\mathbb{P}$ can be used to establish
the reflection, transmission, and spatial field pattern for the structure
in question. The details are given in \cite{BBL}. 

Note that the form of (\ref{eq:abcd_n}--\ref{eq:v1234_n}) is quite
general, and can be simplified greatly in some cases. For instance,
for non-gyrotropic isotropic media ($\alpha=\beta=0$, $\varepsilon$~and~$\mu$
are scalar) after some algebra one can reduce (\ref{eq:abcd_n}) to

\begin{equation}
\mathbb{M}^{(i)}=\left[\begin{array}{cc}
0 & \epsilon I\\
\mu I & 0\end{array}\right].\label{eq:m_iso}\end{equation}

Note that we have used the symbol~$\epsilon$ rather than~$\varepsilon$
to distinguish between scalar and tensor permittivity, respectively.
To avoid confusion, this will be done throughout the article. In what
follows, the rest quantities ($\mu$,~$\alpha$,~$\beta$) will
be assumed scalar unless explicitly stated otherwise.

As a final note, one can see from~(\ref{eq:diff}) that the block
vector $\mathbb{W}$ contains only in-plane components of the field
vectors. Therefore in the case of normal incidence and isotropic ambient
medium with refractive index~$n_{0}$\begin{eqnarray}
\mathbb{W}=\left[\begin{array}{c}
I\mathbf{H}\\
\mathbf{q}^{\times}\mathbf{E}\end{array}\right]=\left[\begin{array}{c}
\mathbf{H}\\
\gamma\mathbf{H}\end{array}\right],\label{eq:impedance}\\
\gamma=\frac{1}{n_{0}}\left(\mathbf{a}_{0}\otimes\mathbf{a}_{0}+\mathbf{b}_{0}\otimes\mathbf{b}_{0}\right)=\frac{1}{n_{0}}I.\nonumber \end{eqnarray}
where~$\gamma$ is called a surface impedance tensor and can be derived
for this case from \cite{BBL}. Note that once we consider finite-sized
multilayers,~$\gamma$ depends \emph{only} on the properties of ambient
medium and thus always has the form~(\ref{eq:impedance}), since
the resulting propagator~(\ref{eq:multipl}) acts in~(\ref{eq:prop})
at the points just outside the structure boundaries.

\section{Multilayer structures with bi-isotropic layers\label{sec:Biisotropic}}

In this section, we consider the simplest case of gyrotropic materials,
when all the four tensors ($\epsilon$,~$\mu$,~$\alpha$,~$\beta$)
are scalar. Such parameters correspond, for instance, to cubic crystals
or optically active liquids, which are sometimes called \emph{bi-isotropic}
or \emph{chiral} media. Because of their relative simplicity, these
media were subject to earlier general analytical studies both as bulk
media \cite{FedorovEng,Lakhtakia88} and as stratified media or multilayers
\cite{Lakhtakia89,JaggardIEEE,JaggardSun,Georgieva,Slepyan,FloodJag,Cory,Becchi,Kim}.
However, as noted above, the latter was either focused on general
problems of a chiral layer interface \cite{Lakhtakia89,JaggardIEEE,JaggardSun,Georgieva}
or concerned with periodic structures \cite{Lakhtakia89,Slepyan,FloodJag,Cory,Becchi}.
Even in the recent, more general theoretical works \cite{Kim} the
authors still confine their numerical studies to periodic media with
bi-isotropic layers.

Similarly to uniaxially birefringent ones, bi-isotropic media are
known to change the effective refractive index depending on the beam
polarization, but with circular rather than linear eigenpolarizations
(see \cite{Georgieva} and references therein). Accordingly, one may
expect that an NPD multilayer with bi-isotropic layers will exhibit
similar polarization-induced splitting \cite{Warsaw} and act as a
polarizer for circularly polarized light.

Numerical experiments, however, show that this is entirely not the
case. We have used the same system as in \cite{Warsaw}, a 27-layer
GaN/air fractal structure shown in the inset of figure~\ref{fig:splitting}.
All GaN layers are artificially rendered isotropic by setting $\epsilon_{o}=\epsilon_{e}$.
Due to absence of birefringence, polarization-induced splitting vanishes,
so only one transmission peak remains and the spectrum becomes that
for the extraordinary wave in figure~\ref{fig:splitting}b. Variable
isotropic optical activity is then introduced by imposing scalar non-zero~$\alpha$
(ranging up to $0.2$, which is somewhat unrealistically high but
chosen so as to fully cover the range currently available in experiments),
so all GaN layers become bi-isotropic. 

\begin{figure}[b]
\begin{center}\includegraphics[%
  width=0.30\linewidth]{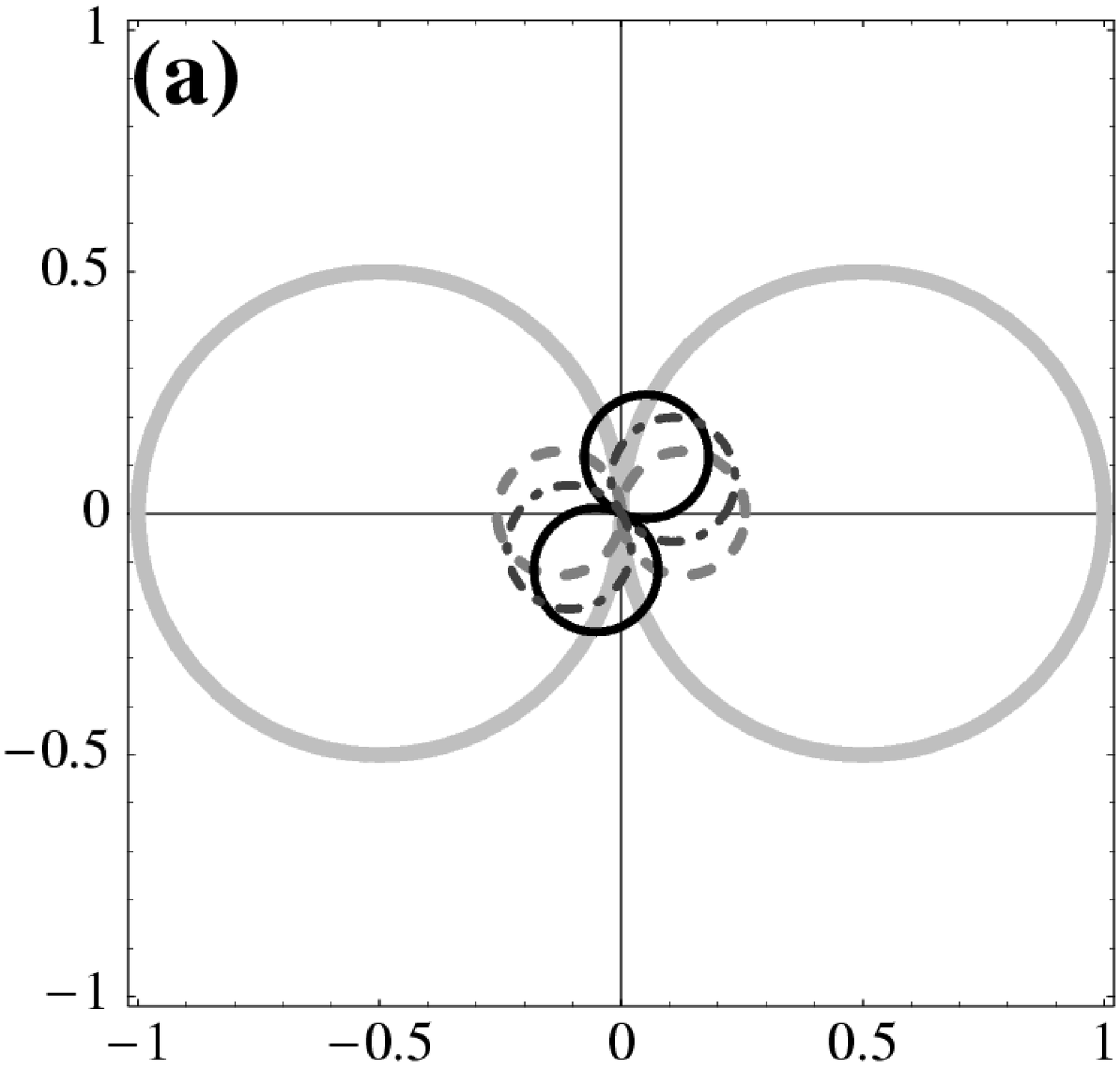}\includegraphics[%
  width=0.30\linewidth]{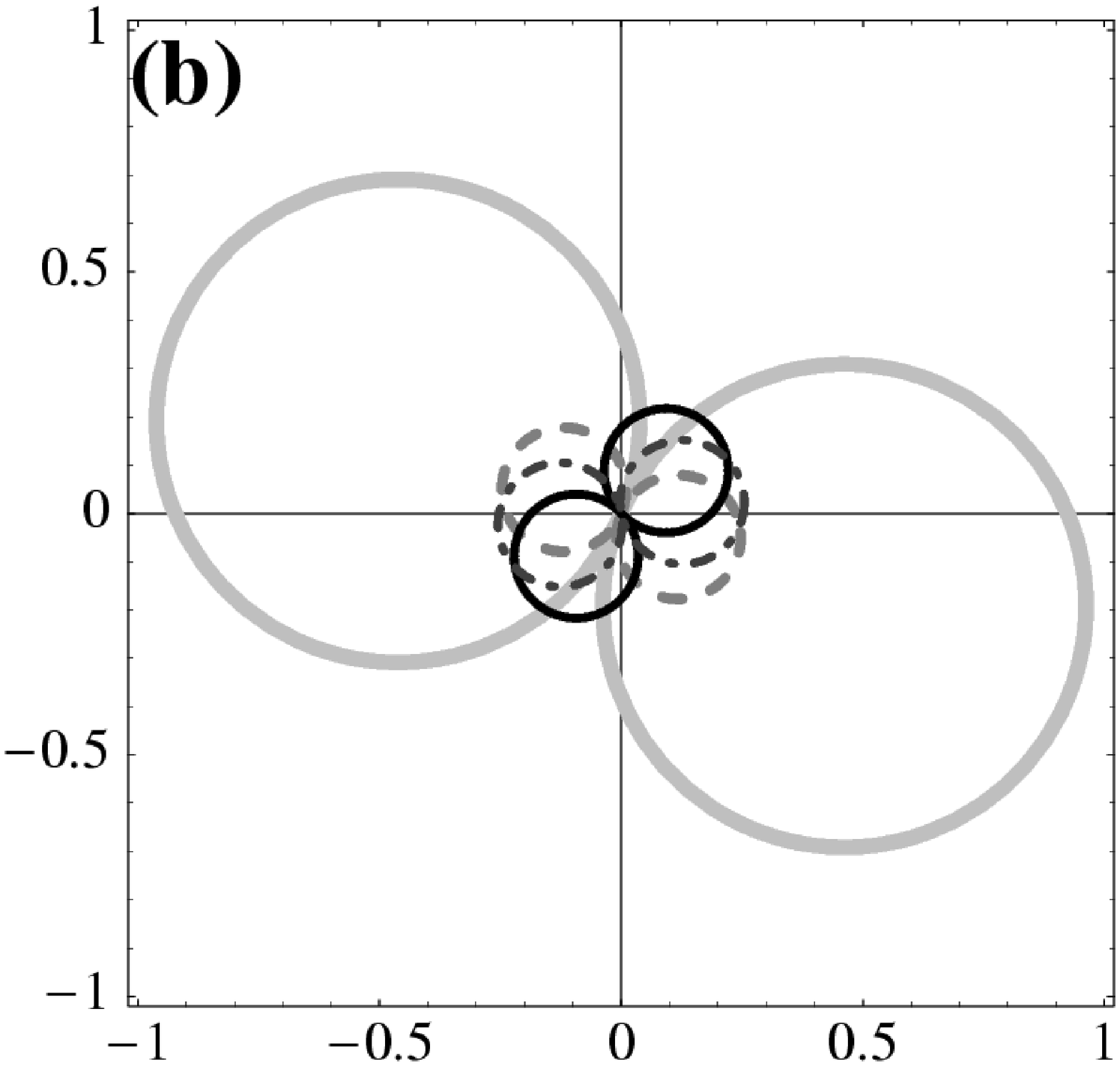}\includegraphics[%
  width=0.30\linewidth]{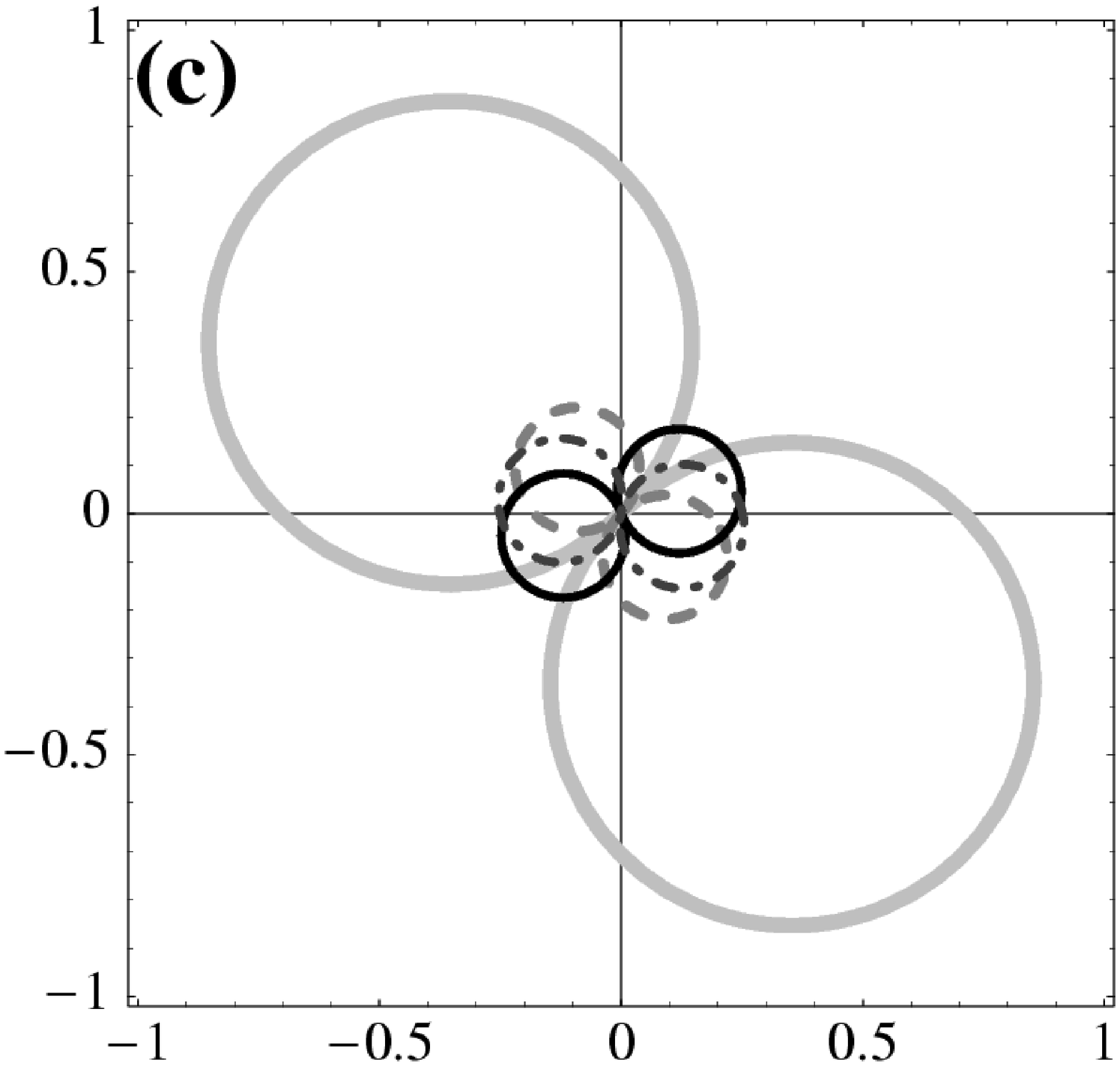}\end{center}

\begin{center}\includegraphics[%
  width=0.30\linewidth]{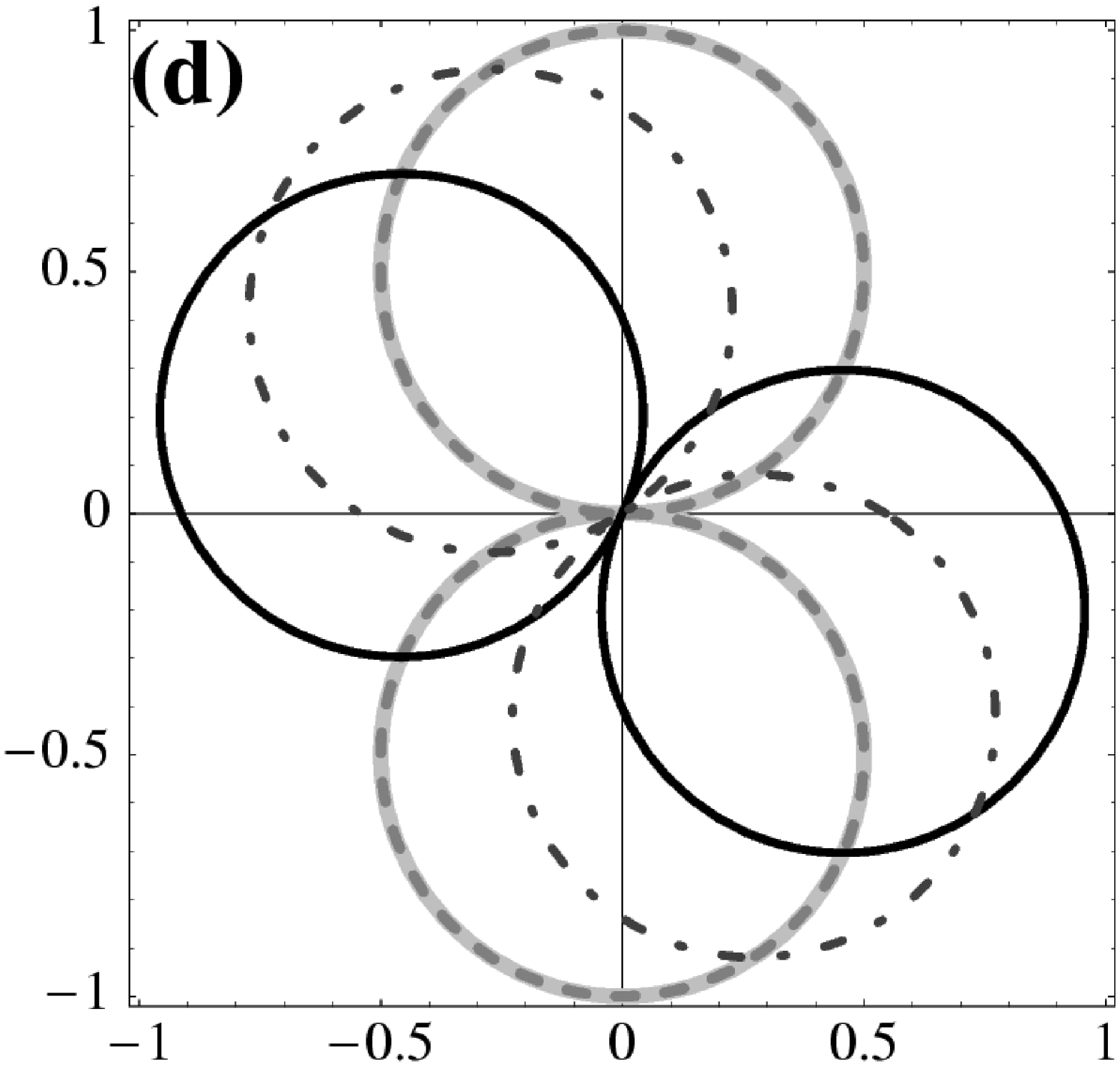}\includegraphics[%
  width=0.30\linewidth]{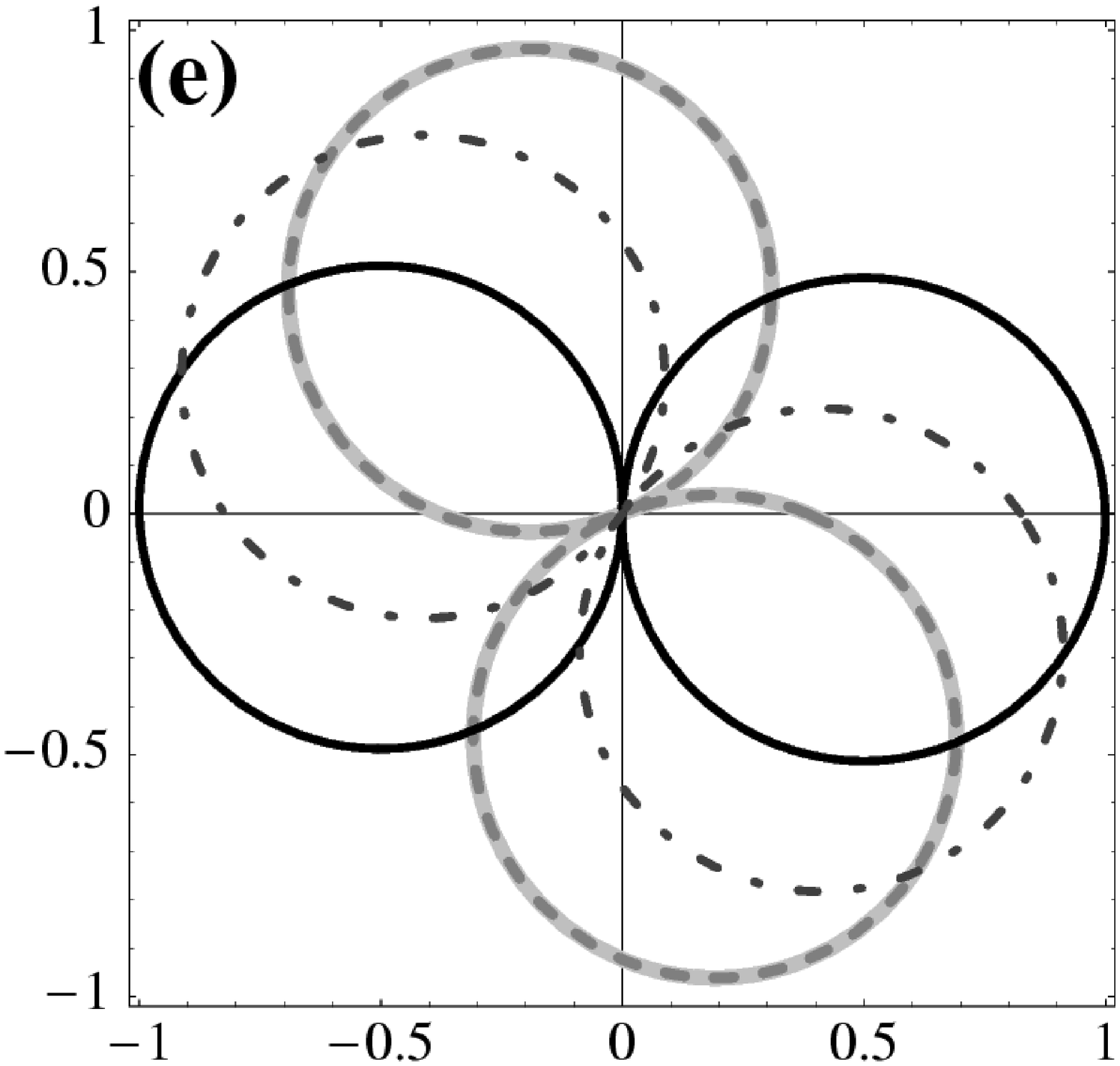}\includegraphics[%
  width=0.30\linewidth]{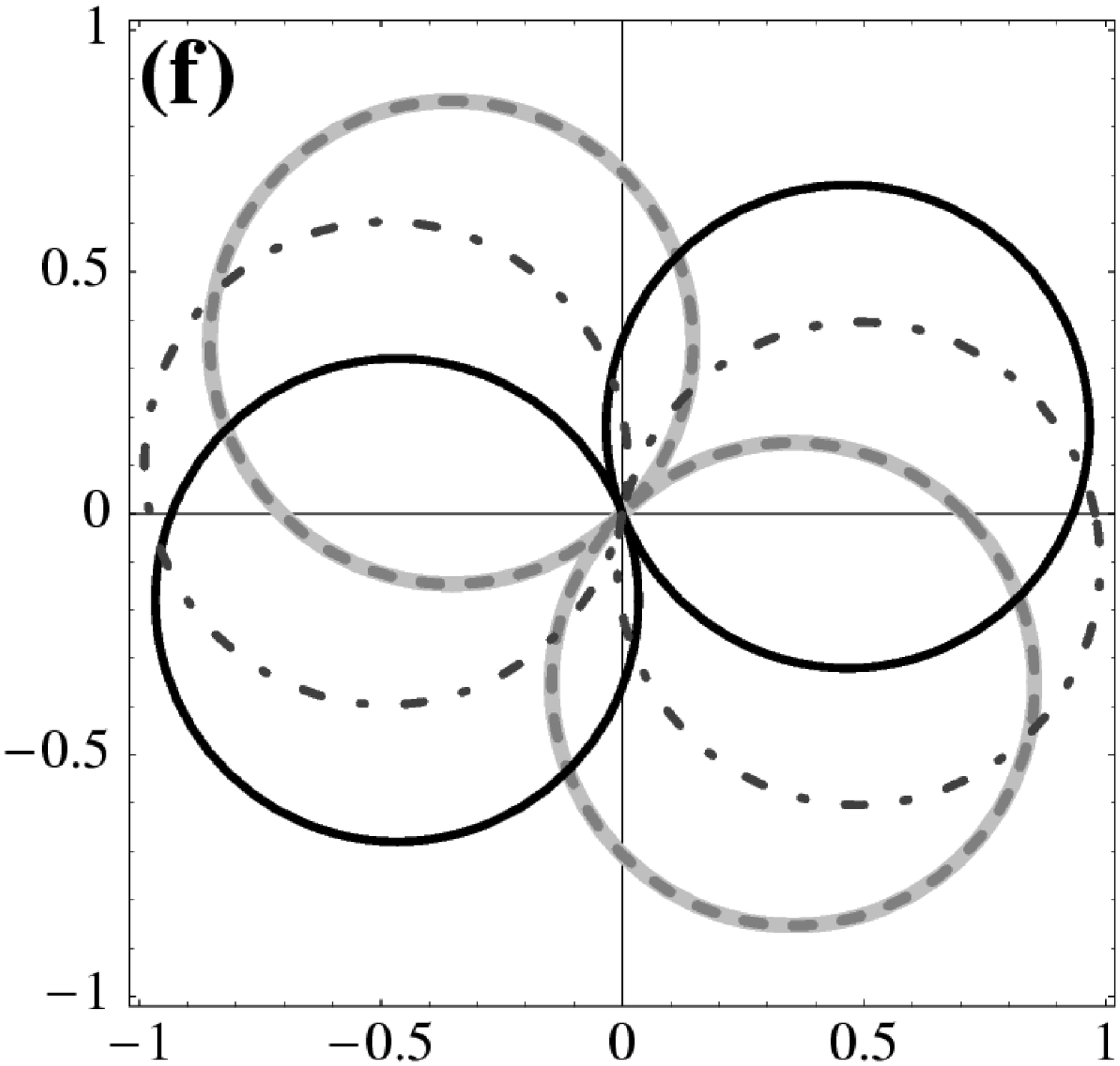}\end{center}

\begin{center}\includegraphics[%
  width=0.70\linewidth]{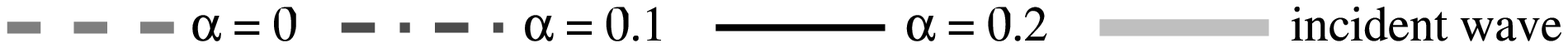}\end{center}

\caption{\label{fig:biiso_output}Polar plots of transmitted wave and incident
wave polarization diagrams (namely, the dependence of $\left|\left(\mathbf{x}\otimes\mathbf{x}\right)\mathbf{H}\right|(\psi)$
where the vector $\mathbf{x}(\psi)=\mathbf{a}_{0}\cos\psi+\mathbf{b}_{0}\sin\psi$
rotates in the plane of interface) for a 27-layer fractal multilayer
containing bi-isotropic layers with variable optical activity~$\alpha$.
The plots are given for off-peak {[}$\omega/\omega_{0}=1.0930$, \textbf{(a-c)}{]}
and on-peak frequencies {[}$\omega/\omega_{0}=1.0938$, \textbf{(d-f)}{]}.
The incident wave polarization varies between $0^{\circ}$~\textbf{(a)},
$22.5^{\circ}$~\textbf{(b)}, $45^{\circ}$~\textbf{(c,f)}, $67.5^{\circ}$~\textbf{(e)},
and $90^{\circ}$~\textbf{(d)}.}
\end{figure}

As~$\alpha$ increases, the transmission spectrum shows no modification
whatsoever. The reflected wave does not change its polarization, either.
The transmitted wave, however, exhibits polarization rotation. Figure~\ref{fig:biiso_output}
shows polarization diagrams for the transmitted wave for different
input polarizations and for different values of~$\alpha$, for two
incident wave frequencies. These polarization diagrams is what a photodetector
would register if placed behind a rotating analyzer in front of the
output beam. It can be seen that this rotation is the same both at
the transmission peak (figure~\ref{fig:biiso_output}d-f) and off-peak
(figure~\ref{fig:biiso_output}a-c). So the polarization rotation
is \emph{uniform}, i.e., it has a similar manner at all frequencies
and regardless of the spectral features. The angle of rotation can
be written as \begin{equation}
\phi=\sum_{j=1}^{N}\frac{\omega}{c}\alpha_{j}d_{j}\label{eq:rotation}\end{equation}

So it can be concluded that the influence of gyrotropy in this case
is \emph{completely incoherent}, i.e., totally independent of the
geometrical properties of the multilayer.

\begin{figure}
\begin{center}\includegraphics[%
  scale=0.3,
  angle=-90]{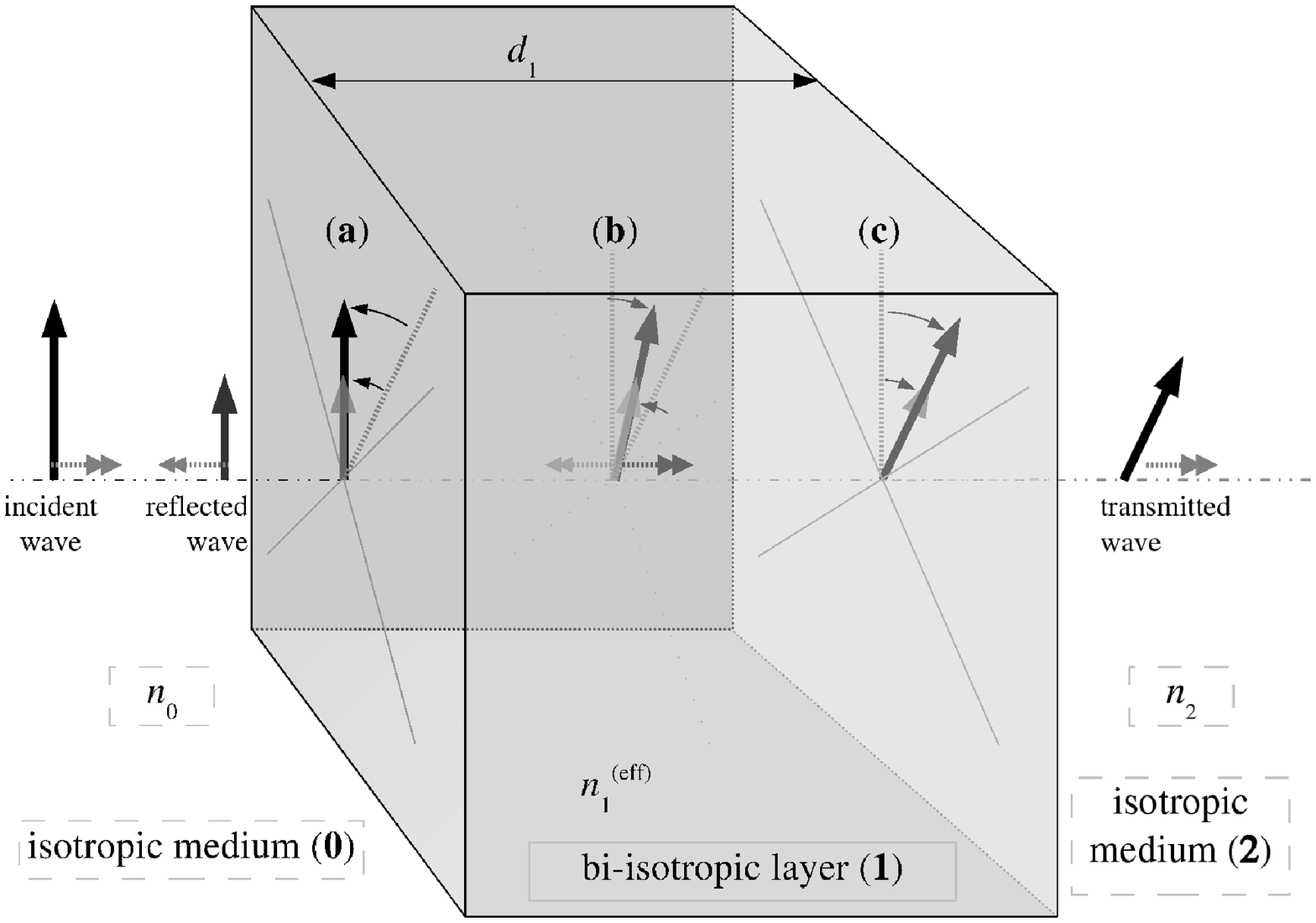}\end{center}

\caption{\label{fig:rotation}The illustration of spectrally uniform polarization
rotation in bi-isotropic media. First, the incident wave hits the
left interface \textbf{(a)}. As the wave passes through the layer,
its polarization is rotated \textbf{(b)}. At the right interface \textbf{(c)}
the wave is partially transmitted and partially back-reflected. As
the back-reflected wave travels back, its polarization is rotated
back \textbf{(b)}, so it arrives at the left interface at exactly
the same polarization as the incident wave \textbf{(a)}. The process
is repeated as this wave is partially reflected and partially transmitted
again. (Double arrows denote propagation direction of the waves shown
by the direction of their $\mathbf{H}$-vectors).}
\end{figure}

To illustrate this seemingly unexpected result, let us recall that
in bi-isotropic media the polarization rotation direction is known
to be reversed after normal-incidence reflection (so-called cross-polarized
case described in \cite{JaggardSun}), in full accordance with the
fact that the medium itself does not exhibit any specific directions.
So the reflected part of the incident beam (see figure~\ref{fig:rotation})
will come to the initial interface ($0|1$) at its incident polarization,
and this is true regardless of the number of multiple reflections
taken into consideration. Similarly, at the second layer interface
($1|2$) all beams will have the same polarization, rotated with respect
to the incident wave polarization according to~(\ref{eq:rotation}).
Since the transmittance is determined by interference effects resulting
from multiple reflections, optical activity in a bi-isotropic layer
cannot make any contribution except the above-mentioned polarization
rotation (figure~\ref{fig:rotation}).

This explanation should hold regardless of the number of layers in
a multilayer structure or of the structure geometry, because interference
of multiply reflected beams should occur likewise at all the layer
interfaces provided all the constituent layers are either isotropic
or bi-isotropic. This was shown for periodic stacks in \cite{FloodJag}
and can also be derived from single-layer and single-interface results
obtained in \cite{Lakhtakia89,JaggardSun}.

To provide a consistent back-up for arbitrary NPD multilayers, let
us calculate the evolution operator for a bi-isotropic layer. Starting
with~(\ref{eq:abcd_n}--\ref{eq:v1234_n}), one can immediately notice
that for all the auxiliary vectors~(\ref{eq:v1234_n}) the expressions
in brackets are scalars and thus commute with $\mathbf{q}$, leaving
the product~$\mathbf{q}I$, which is identically zero. Remembering
that $I\mathbf{q}^{\times}=\mathbf{q}^{\times}I=\mathbf{q}^{\times}$
and that the absence of absorption implies $\beta=-\alpha$, (\ref{eq:abcd_n})~can
be reduced to

\begin{equation}
\mathbb{M}^{(b)}=\left[\begin{array}{cc}
i\alpha\mathbf{q}^{\times} & \epsilon I\\
\mu I & i\alpha\mathbf{q}^{\times}\end{array}\right].\label{eq:m_biiso}\end{equation}

Taking the matrix exponential, one can, after some algebra, find the
propagator to equal %
\footnote{If we are working in a 3D real space, the propagator will formally
contain an additional term $\mathbf{q}\otimes\mathbf{q}$ added to
diagonal terms. However, in our system only normal incidence is considered.
Hence for all field vectors $\mathbf{qH}=0$. Since, naturally, $I\mathbf{q}=0=\mathbf{q}^{\times}\mathbf{q}$,
all calculations are unchanged if performed in a 2D subspace associated
with the interface plane. Formally speaking, we can consider blocks
of~$\mathbb{M}$ to be~$2\times2$ rather than~$3\times3$ matrices
before taking the exponential.%
}

\begin{equation}
\begin{array}{c}
\mathbb{P}^{(b)}=\left[\begin{array}{cc}
\cos(kd\sqrt{\epsilon\mu})P(\alpha) & \frac{i\epsilon}{\sqrt{\epsilon\mu}}\sin(kd\sqrt{\epsilon\mu})P(\alpha)\\
\frac{i\mu}{\sqrt{\epsilon\mu}}\sin(kd\sqrt{\epsilon\mu})P(\alpha) & \cos(kd\sqrt{\epsilon\mu})P(\alpha)\end{array}\right]\\
P(\alpha)\equiv\left[\cos(kd\alpha)I-\sin(kd\alpha)\mathbf{q}^{\times}\right]\end{array}\label{eq:p_biiso}\end{equation}
where~$k\equiv\omega/c$.

The evolution operator for the isotropic layer with the same~$\epsilon$,~$\mu$,~and~$d$
can be found either by imposing $\alpha=0$ or directly from~(\ref{eq:m_iso}),
which results in

\begin{equation}
\mathbb{P}^{(i)}=\left[\begin{array}{cc}
\cos(kd\sqrt{\epsilon\mu})I & \frac{i\epsilon}{\sqrt{\epsilon\mu}}\sin(kd\sqrt{\epsilon\mu})I\\
\frac{i\mu}{\sqrt{\epsilon\mu}}\sin(kd\sqrt{\epsilon\mu})I & \cos(kd\sqrt{\epsilon\mu})I\end{array}\right].\label{eq:p_iso}\end{equation}

We can see that every real-space block of both~(\ref{eq:p_biiso})
and~(\ref{eq:p_iso}) has identical geometrical structure, with different
coefficients for different components of the block vectors~$\mathbb{W}$
in~(\ref{eq:prop}). This can be written as\begin{equation}
\mathbb{P}^{(b)}=\left(\cos(kd\alpha)I-\sin(kd\alpha)\mathbf{q}^{\times}\right)\mathbb{L},\quad\mathbb{P}^{(i)}=I\mathbb{L}\label{eq:p_short}\end{equation}
where\begin{equation}
\mathbb{L}=\left[\begin{array}{cc}
\cos(kd\sqrt{\epsilon\mu}) & \frac{i\epsilon}{\sqrt{\epsilon\mu}}\sin(kd\sqrt{\epsilon\mu})\\
\frac{i\mu}{\sqrt{\epsilon\mu}}\sin(kd\sqrt{\epsilon\mu}) & \cos(kd\sqrt{\epsilon\mu})\end{array}\right].\label{eq:lambda}\end{equation}

Here one can note that the block matrix~$\mathbb{L}$ consists of
scalar quantities, so in terms of real space it is a coefficient matrix,
which does not affect polarization properties. This essentially results
from normal incidence. Also note that this coefficient matrix is the
only part in~$\mathbb{P}$ that depends on propagation phase $\varphi=\sqrt{\epsilon\mu}d\omega/c$. 

Now let us show what such a propagator does to a normally incident
vector~$\mathbf{H}_{0}$. For simplicity and without loss of generality,
we can assume $\mathbf{H}_{0}=H_{0}\mathbf{a}_{0}$, after which it
can be seen from~(\ref{eq:impedance}) that propagation in isotropic
layer results in \begin{equation}
\mathbb{W}=\mathbb{P}^{(i)}\mathbb{W}_{0}=H_{0}I\mathbb{L}\left[\begin{array}{c}
\mathbf{a}_{0}\\
\frac{1}{n_{0}}\mathbf{a}_{0}\end{array}\right]=H_{0}\mathbb{L}\left[\begin{array}{c}
1\\
\frac{1}{n_{0}}\end{array}\right]I\mathbf{a}_{0}=H_{0}\left[\begin{array}{c}
t\\
t'\end{array}\right]\mathbf{a}_{0}.\label{eq:w_iso}\end{equation}

One can see that there is no polarization rotation. On the other hand,
substituting the evolution operator of a bi-isotropic layer analogously
yields

\begin{equation}
\mathbb{W}=\mathbb{P}^{(b)}\mathbb{W}_{0}=H_{0}\left[\begin{array}{c}
t\\
t'\end{array}\right]\left(\mathbf{a}_{0}\cos(kd\alpha)-\mathbf{b}_{0}\sin(kd\alpha)\right).\label{eq:w_biiso}\end{equation}

We can see that the polarization of the transmitted wave is rotated
by the angle of $\phi=kd\alpha$, while the values of~$t$~and~$t'$
(which contribute to the transmittance) remain the same. Besides,
it can be stated that the propagator structure for the bi-isotropic
layer in~(\ref{eq:p_short}) explicitly contains the value of~$\phi$.

To advance from a single layer to a multilayer structure, let us investigate
what happens when propagators in the form~(\ref{eq:p_short}) are
multiplied. Naturally, we assume different $\varepsilon$,~$\mu,$~$d$,
and~$\alpha$ (if any) for the two adjacent layers.

So, a product of two isotropic propagators reads\begin{equation}
\mathbb{P}_{12}^{(ii)}=\mathbb{P}_{1}^{(i)}\mathbb{P}_{2}^{(i)}=\left(\mathbb{L}_{1}I\right)\left(\mathbb{L}_{2}I\right)=\left(\mathbb{L}_{1}\mathbb{L}_{2}\right)I=\mathbb{L}_{12}I\label{eq:mult_ii}\end{equation}

One can see that while the coefficient matrices~$\mathbb{L}_{j}$
are multiplied (and hence the complex \emph{spectral} effects are
seen in isotropic multilayer structures), the real-space structure
of the propagators remains proportional to~$I$, i.e., \emph{polarization-independent}. 

Likewise, a product between isotropic and bi-isotropic propagators
reads\begin{equation}
\mathbb{P}_{12}^{(ib)}=\mathbb{P}_{1}^{(i)}\mathbb{P}_{2}^{(b)}=\mathbb{L}_{12}\left(\cos(kd_{2}\alpha_{2})I-\sin(kd_{2}\alpha_{2})\mathbf{q}^{\times}\right)\label{eq:mult_ib}\end{equation}

Again, the coefficient matrices are multiplied in the same manner,
while real-space geometrical structure is preserved, so the angle
of polarization rotation is unchanged compared to that for a single
bi-isotropic layer.

Finally, taking a product between two bi-isotropic propagators, after
some algebra we arrive at\begin{equation}
\mathbb{P}_{12}^{(bb)}=\mathbb{L}_{12}\left(\cos(kd_{1}\alpha_{1}+kd_{2}\alpha_{2})I-\sin(kd_{1}\alpha_{1}+kd_{2}\alpha_{2})\mathbf{q}^{\times}\right).\label{eq:mult_bb}\end{equation}

So we see that the coefficient matrices are again multiplied likewise,
and hence no change to the spectral properties provided that the same~$\mathbb{L}_{1}$
and~$\mathbb{L}_{2}$ are used in~(\ref{eq:mult_ii})~and~(\ref{eq:mult_bb})
-- which is true if the corresponding layers have the same $\varepsilon$,~$\mu$,~and~$d$.
Looking at the geometrical structure of the propagators, we see that
\begin{equation}
\phi_{12}=\phi_{1}+\phi_{2}=\frac{\omega}{c}\left(\alpha_{1}d_{1}+\alpha_{2}d_{2}\right).\label{eq:phi_bb}\end{equation}

Certainly, as the propagator for any multilayer is built according
to~(\ref{eq:multipl}), and any multiplication according to~(\ref{eq:mult_ii}--\ref{eq:mult_bb})
leads to the fact that spectra are independent of~$\alpha$ while
polarization rotation sums up, so~(\ref{eq:phi_bb}) is directly
generalized to~(\ref{eq:rotation}). Therefore the reasoning applicable
to a single bi-isotropic layer holds for any fractal structure (or,
for that matter, for an arbitrarily designed binary multilayer composed
of isotropic and bi-isotropic layers).

So far we have considered a linearly polarized incident wave. If the
polarization is circular, it can be shown that change in effective
refractive index does occur, however it does not manifest itself in
the transmission spectra. Detailed explanation is provided in~\ref{sec:app_a}.

To summarize, we have shown analytically that in multilayers consisting
of isotropic and bi-isotropic layers in arbitrary combination \emph{the
optical spectra are exactly the same as if there were no optical activity
whatsoever}, regardless of incident wave polarization. Instead, optical
activity rotates the transmitted wave polarization, and the amount
of this rotation is totally independent of the multilayer's geometrical
composition and is described by a simple equation~(\ref{eq:rotation}).

This result can also be understood from the point of view that in
a bi-isotropic multilayer system the multiple-reflected beams will
interfere at each layer interface in exactly the same way as they
do in the isotropic multilayer \emph{}with the same $\epsilon$,~$\mu$,~and~$d$.
Since the optical spectra are determined by the nature of such interference,
they are unmodified despite polarization rotation caused by each bi-isotropic
layer and resulting in overall uniform polarization rotation.

Note that the analytical derivation in this case is rigorous and exact,
without any assumptions on the values of parameters used. The only
assumption regards the general applicability of local material equations~(\ref{eq:mat}),
which restricts the layer thickness to the wavelength such that $d\gg0.001\lambda$
\cite{Fedorov} -- a condition well met in the current state-of-the-art
multilayers in the optical range.

In conclusion to this section, let us point out that the uniform and
incoherent nature of the polarization plane rotation allows to use
bi-isotropic materials to rotate polarization in devices with any
geometry, without risk of additional interference effects. It can
also be possible to combine spectral and polarization-rotating device
in one multilayer, which should be beneficial for miniaturization
of integrated optical components.

\section{Gyrotropy and polarization-induced splitting\label{sec:Gyrotropy}}

\begin{figure}
\begin{center}\includegraphics[%
  bb=0bp 25bp 845bp 321bp,
  width=1.0\linewidth]{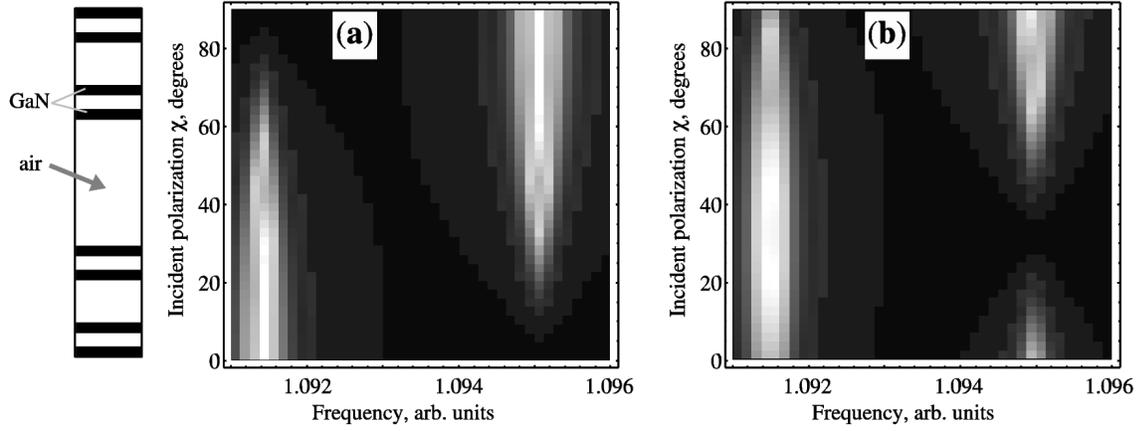}\end{center}

\caption{\label{fig:gyro_split}The dependence of transmittance on frequency
and incident wave polarization for a 27-layer GaN/air fractal structure
(\emph{inset}) without \textbf{(a)} and with gyrotropy \textbf{(b)}
with $\alpha=0.2$. Lighter areas denote higher transmittance. }
\end{figure}

Gradually complicating our structure, let us now consider a birefringent
medium with a uniaxial permittivity tensor \begin{equation}
\varepsilon=\epsilon_{o}+\left(\epsilon_{e}-\epsilon_{o}\right)\mathbf{c}\otimes\mathbf{c}\label{eq:eps_uniaxial}\end{equation}
 where, as noted before,~$\varepsilon$~and~$\epsilon$ are used
to discriminate between tensor and scalar permittivity, respectively,
and subscripts~$o$ and~$e$ stand for {}``ordinary'' and {}``extraordinary''.
A unit vector~$\mathbf{c}$ determines the optical axis orientation.
In multilayer structures under study, all uniaxial layers have similar
orientation of~$\mathbf{c}$ such that $\mathbf{cq}=0$. Without
loss of generality we can assume $\mathbf{c}=\mathbf{b}_{0}$. As
before, we assume no magnetic anisotropy, and isotropic optical activity. 

Substituting~(\ref{eq:eps_uniaxial}) into~(\ref{eq:abcd_n}--\ref{eq:v1234_n})
lets us arrive after some simple algebra at\begin{equation}
\mathbb{M}^{(g)}=\left[\begin{array}{cc}
i\alpha\mathbf{q}^{\times} & \varepsilon I+iQ\mu_{q}\mathbf{q}^{\times}\varepsilon\mathbf{q}\otimes\mathbf{q}^{\times}\mathbf{q}\varepsilon I\\
\mu I & i\alpha\mathbf{q}^{\times}\end{array}\right]=\left[\begin{array}{cc}
i\alpha\mathbf{q}^{\times} & \varepsilon I\\
\mu I & i\alpha\mathbf{q}^{\times}\end{array}\right].\label{eq:m_gyro}\end{equation}

Without optical activity ($\alpha=0$) and within our assumptions
($\mathbf{c}=\mathbf{b}_{0}$) the propagator can be found to equal

\begin{equation}
\mathbb{P}^{(a)}=\mathbb{L}_{o}\mathbf{a}_{0}\otimes\mathbf{a}_{0}+\mathbb{L}_{e}\mathbf{b}_{0}\otimes\mathbf{b}_{0}\label{eq:p_uniaxial}\end{equation}
where

\begin{eqnarray}
\mathbb{L}_{j} & = & \left[\begin{array}{cc}
\cos(kd\sqrt{\epsilon_{j}\mu}) & \frac{i\epsilon_{j}}{\sqrt{\epsilon_{J}\mu}}\sin(kd\sqrt{\epsilon_{j}\mu})\\
\frac{i\mu}{\sqrt{\epsilon_{j}\mu}}\sin(kd\sqrt{\epsilon_{j}\mu}) & \cos(kd\sqrt{\epsilon_{j}\mu})\end{array}\right],\quad j=o,e.\label{eq:lambda_uniax}\end{eqnarray}

It can be easily seen that such a layer is able to exhibit polarization-induced
peak splitting in the spectrum. Indeed, when a field vector is oriented
along~$\mathbf{a}_{0}$ (or, more generally speaking, perpendicular
to~$\mathbf{c}$), it will only interact with one term in the propagator,
namely, the one containing~$\mathbb{L}_{o}$, and will propagate
exactly as if all uniaxial layers were isotropic with $\epsilon=\epsilon_{o}$
{[}compare (\ref{eq:lambda})~and~(\ref{eq:lambda_uniax}){]}. Similarly,
when a field vector is oriented along~$\mathbf{b}_{0}$ (parallel
to~$\mathbf{c}$), the same would be true for~$\mathbb{L}_{e}$,
and again all uniaxial layers will effectively be isotropic, this
time with $\epsilon=\epsilon_{e}$. So, one can see that if the incident
wave is polarized along either of the eigenvectors of~$\mathbb{P}$
(it is then called an \emph{eigenwave}), it will propagate in exactly
the same way as if all uniaxial layers were isotropic with some effective
refractive index. Hence the spectra will be similar in shape, the
only difference associated with a change in effective~$\epsilon$,
which in turn causes all spectral features to shift in frequency.

Unfortunately, in presence of optical activity the tensorial nature
of~$\varepsilon$ results in a very complicated structure of the
propagator, which, while capable of being evaluated symbolically,
is nearly useless for further analysis.

It can be shown, however, that if we assume both~$\alpha$ and~$\delta\epsilon\equiv\frac{1}{2}\left|\epsilon_{e}-\epsilon_{o}\right|$
to be small, an approximate \emph{model propagator} can be used. The
details on its derivation and applicability are covered in ~\ref{sec:app_b}.
Reduced with respect to~$\delta\epsilon$ with the product~$\alpha\delta\epsilon$
neglected, it reads \begin{equation}
\mathbb{P}^{(g)}\approx\mathbb{L}\left(\cos(kd\alpha)I-\sin(kd\alpha)\mathbf{q}^{\times}\right)+\delta\epsilon\mathbb{L}'\left(\mathbf{a}_{0}\otimes\mathbf{a}_{0}-\mathbf{b}_{0}\otimes\mathbf{b}_{0}\right).\label{eq:p_gyro_1}\end{equation}

Here the coefficient matrix~$\mathbb{L}$ is the same as that for
isotropic media {[}compare (\ref{eq:ap_gyro_matrix_1})~and~(\ref{eq:lambda}){]}
if we substitute $\epsilon\equiv\frac{1}{2}\left(\epsilon_{e}+\epsilon_{o}\right)$,
i.e., if an average between ordinary and extraordinary dielectric
constant is used as an effective value of~$\epsilon$. The coefficient
matrix~$\mathbb{L}'$ is explicitly written as~(\ref{eq:ap_gyro_matrix_2}).

On the other hand, one can reduce the same model propagator with respect
to~$\alpha$, in which case\begin{equation}
\mathbb{P}^{(g)}\approx\mathbb{L}_{o}\mathbf{a}_{0}\otimes\mathbf{a}_{0}+\mathbb{L}_{e}\mathbf{b}_{0}\otimes\mathbf{b}_{0}-kd\alpha\mathbb{L}\mathbf{q}^{\times},\label{eq:p_gyro_2}\end{equation}
where the matrices~$\mathbb{L}_{o}$ and~$\mathbb{L}_{e}$ are the
same as those without gyrotropy {[}see~(\ref{eq:p_uniaxial}--\ref{eq:lambda_uniax}){]},
and the additional coefficient matrix~$\mathbb{L}$ given in~(\ref{eq:ap_gyro_matrix_1})
is the same as for isotropic case in~(\ref{eq:lambda}). 

Looking at~(\ref{eq:p_gyro_1})~and~(\ref{eq:p_gyro_2}), one can
see that even within the bounds of the approximation used, the behavior
of the multilayer in question changes dramatically. Compared to a
bi-isotropic evolution operator~(\ref{eq:p_short}), there is a symmetric
addition proportional to~$\delta\epsilon$, which causes perturbation
to polarization plane rotation, making the propagator polarization-sensitive.
Hence, the propagator multiplication is no longer described by~(\ref{eq:mult_bb}),
and the angle of rotation becomes dependent on the incident wave polarization.

Compared to a birefringent evolution operator~(\ref{eq:p_uniaxial}),
there is an antisymmetric addition proportional to~$\alpha$ (and,
more generally, this~$\alpha$ comes from decomposition of~$\sin kd\alpha$).
We know from~(\ref{eq:p_short}) that such a term in the propagator
is responsible for polarization plane rotation. 

As it is not easy to decide in favor of one on these two interpretations,
it is safe to assume that one layer made of material in question will
exhibit polarization rotation combined (more or less on equal terms)
with spectral properties of birefringent multilayer structures.

Numerical results show that this is indeed the case. For simulation,
the same GaN/air fractal multilayer as in the previous section was
used (shown again in figure~\ref{fig:gyro_split}), but without rendering
the GaN layers isotropic. Optical activity was likewise artificially
introduced into GaN, $\alpha$ ranging up to $0.2$. 

\begin{figure}[b]
\begin{center}\includegraphics[%
  width=0.30\linewidth]{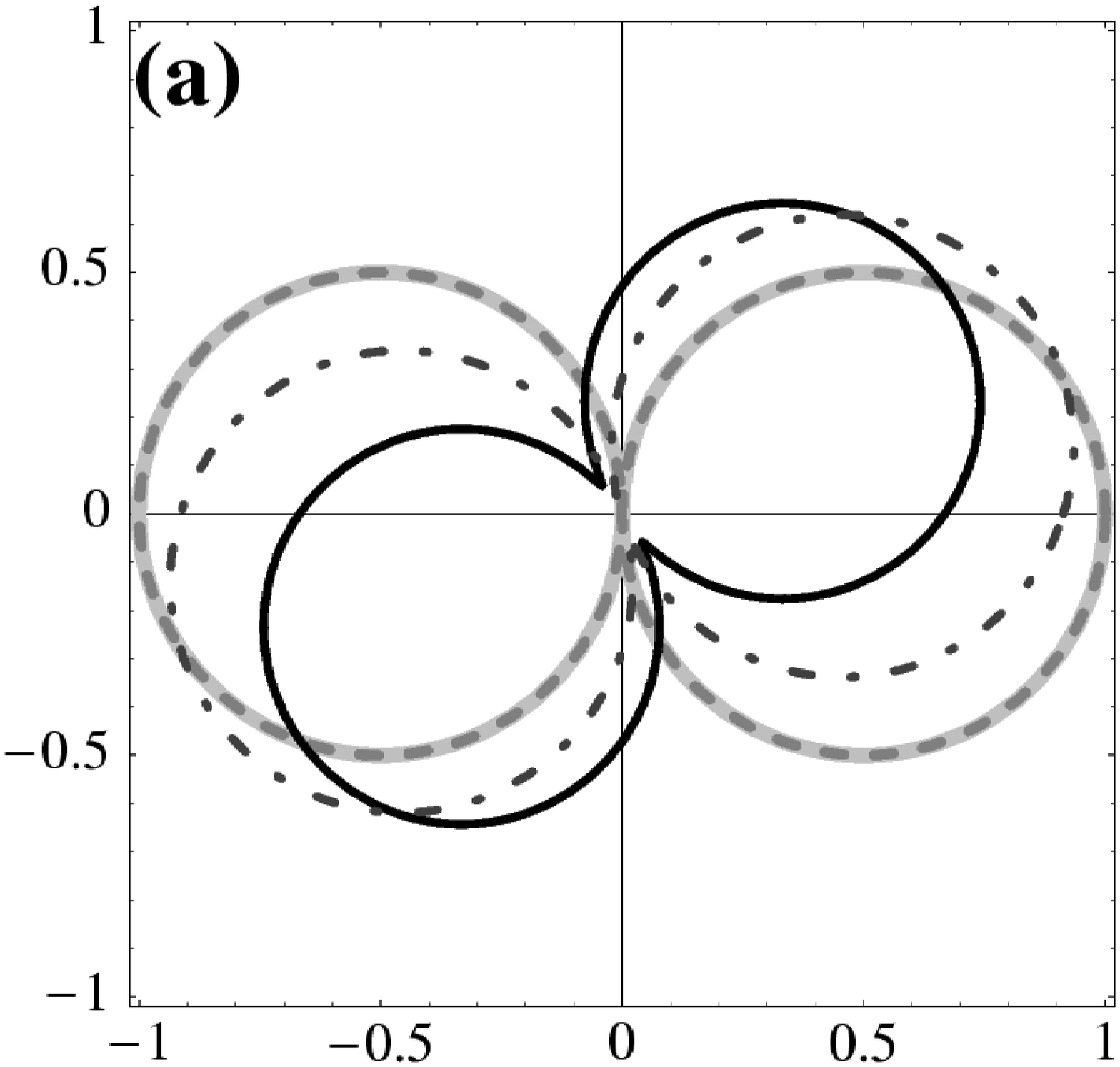}\includegraphics[%
  width=0.30\linewidth]{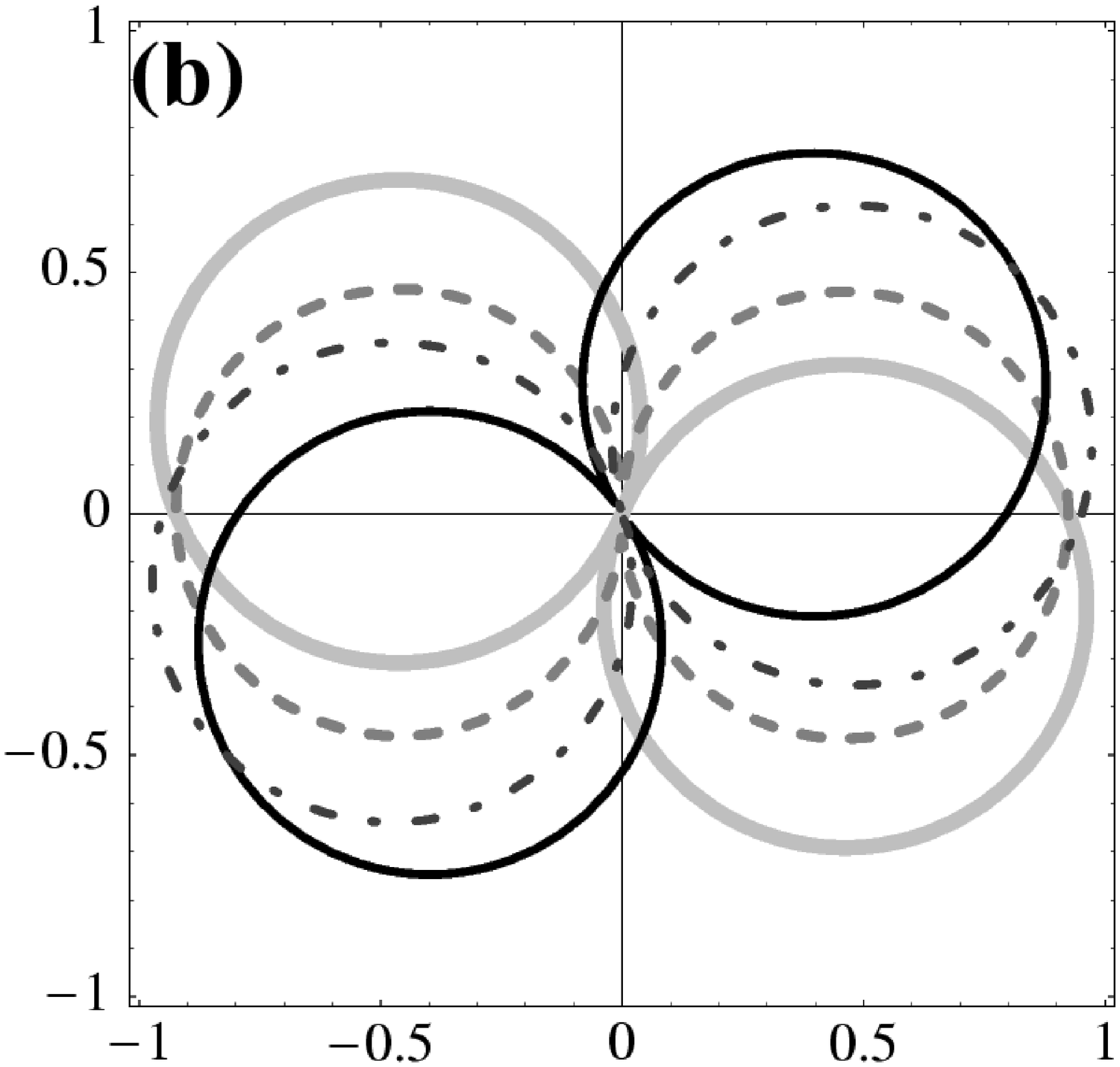}\includegraphics[%
  width=0.30\linewidth]{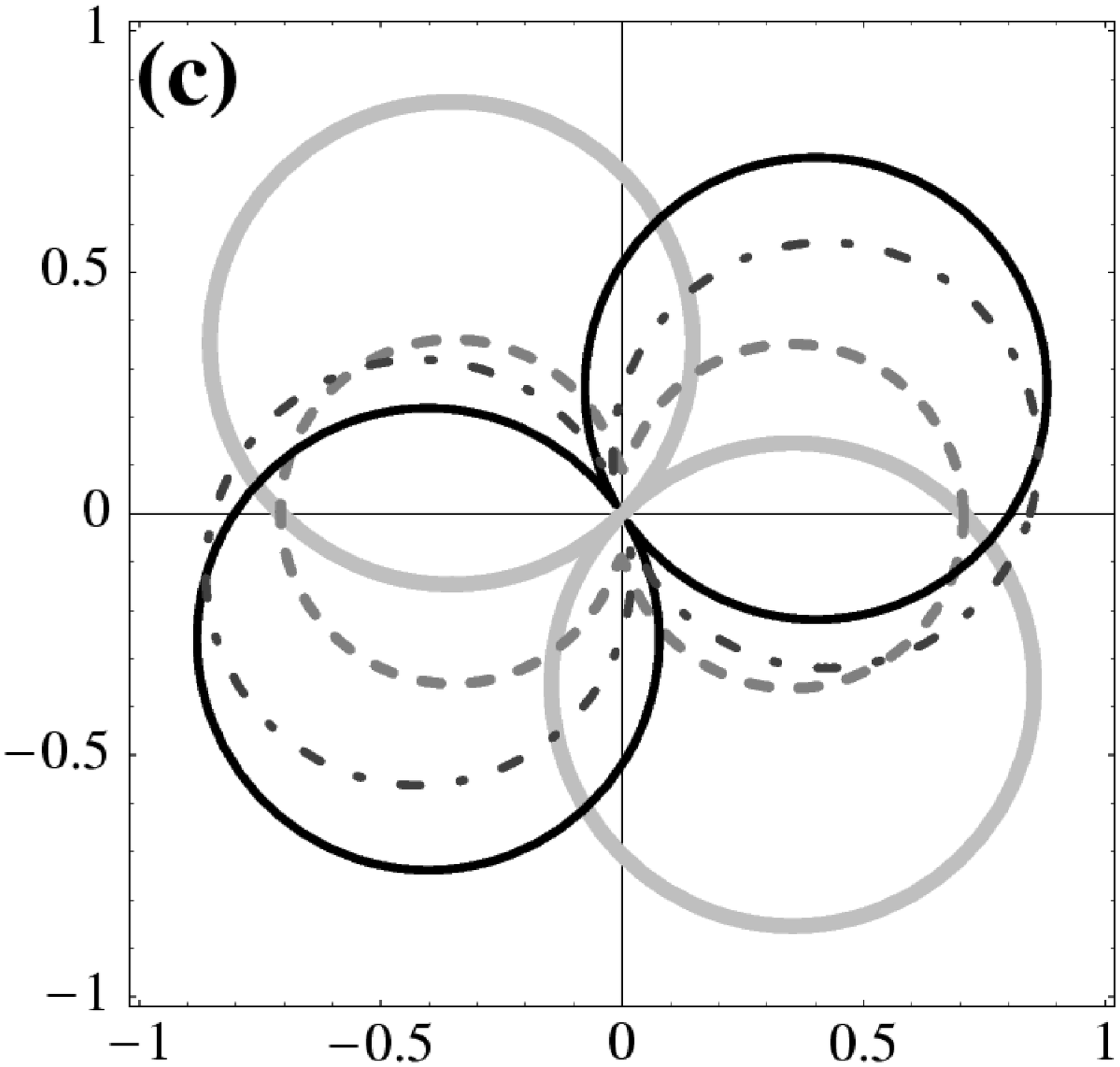}\end{center}

\begin{center}\includegraphics[%
  width=0.30\linewidth]{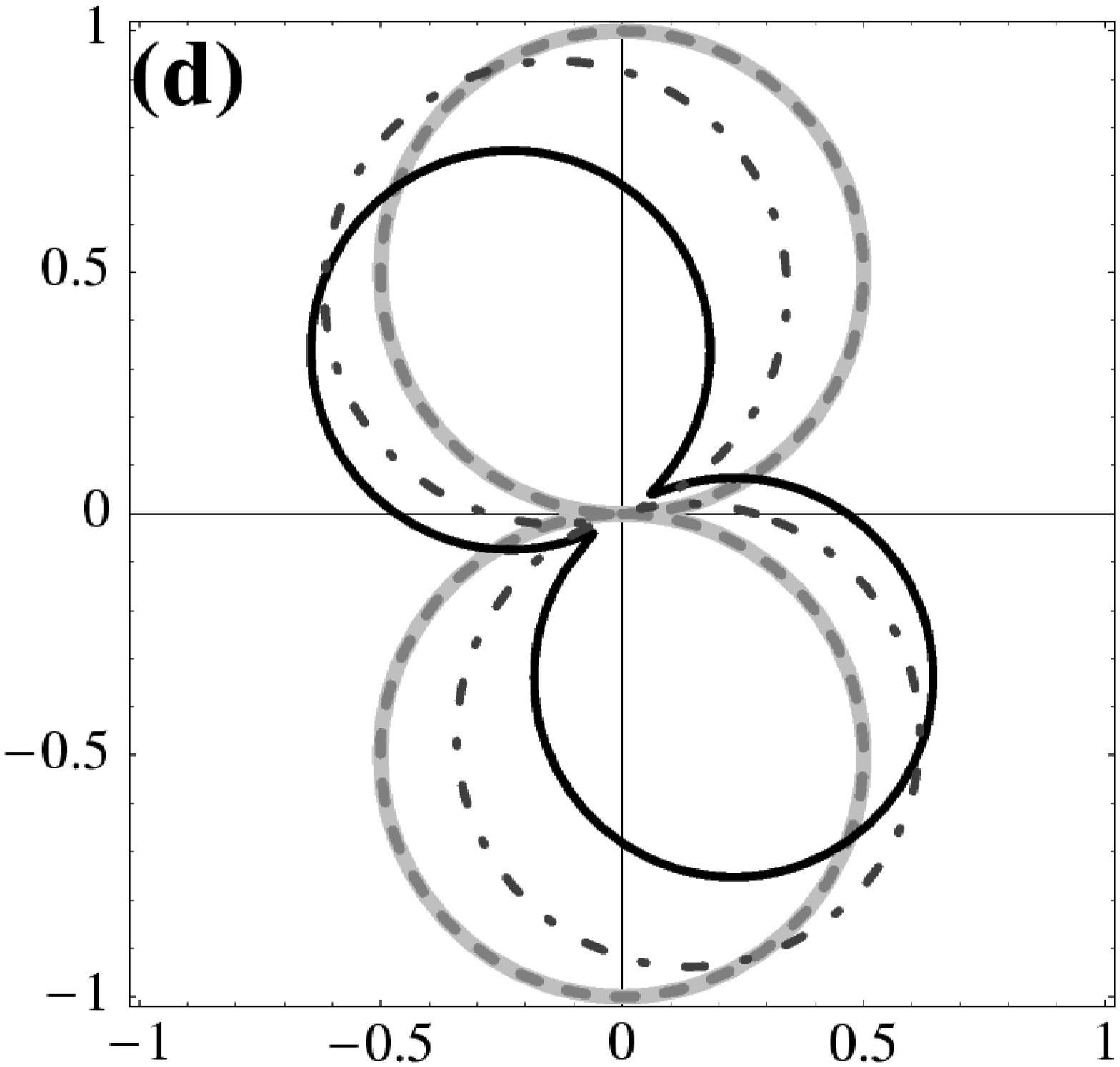}\includegraphics[%
  width=0.30\linewidth]{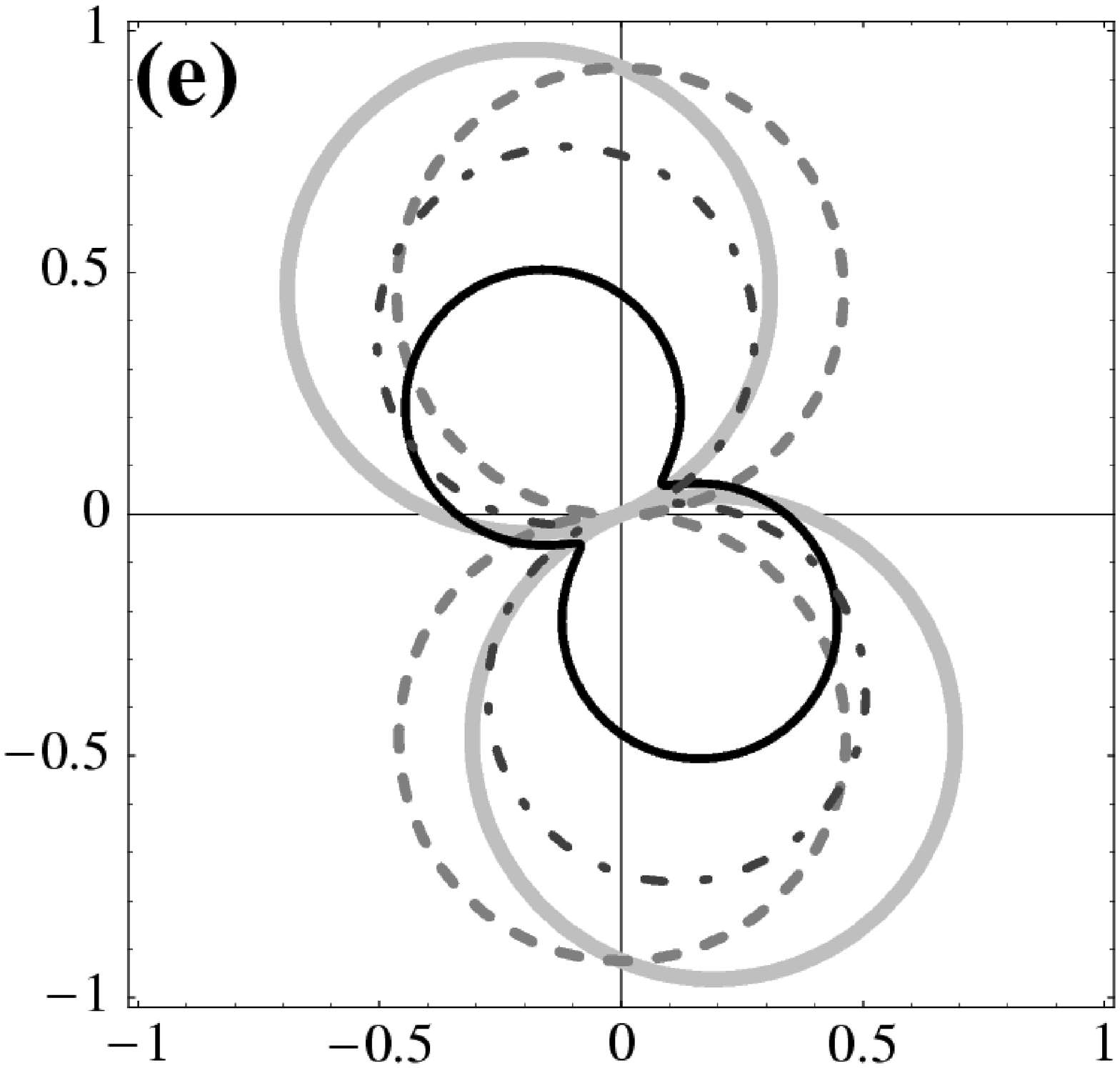}\includegraphics[%
  width=0.30\linewidth]{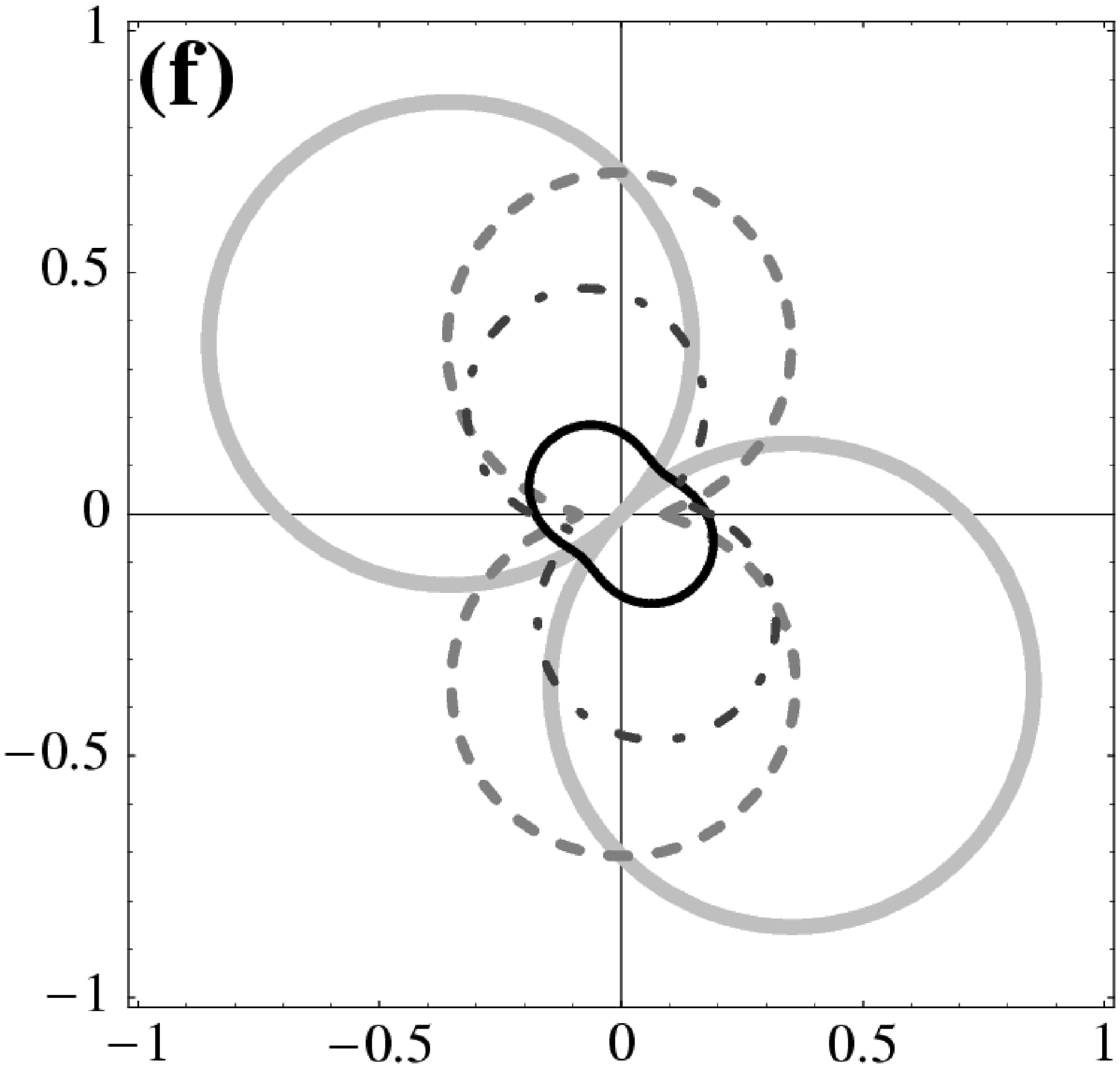}\end{center}

\begin{center}\includegraphics[%
  width=0.70\linewidth]{fig/Fig36aux.eps}\end{center}

\caption{\label{fig:gyro_output}Polar plots of transmitted wave and incident
wave polarization diagrams (the same as in Fig.~\ref{fig:biiso_output})
for a 27-layer fractal multilayer containing birefringent layers with
variable isotropic optical activity~$\alpha$. The plots are given
for the same incident wave polarizations for both polarization-split
peaks {[}$\omega/\omega_{0}=1.0923$ \textbf{(a-c)}, $\omega/\omega_{0}=1.0938$
\textbf{(d-f)}{]}. The incident wave polarization varies between $0^{\circ}$~\textbf{(a)},
$22.5^{\circ}$~\textbf{(b)}, $45^{\circ}$~\textbf{(c,f)}, $67.5^{\circ}$~\textbf{(e)},
and $90^{\circ}$~\textbf{(d)}.}
\end{figure}

As one can see in figure~\ref{fig:gyro_split}, there is no change
in the nature of polarization-induced splitting, and the frequencies
of the doublet components are not changed, either. This agrees with~(\ref{eq:p_gyro_2})
where we see the coefficient matrices~$\mathbb{L}_{o}$~and~$\mathbb{L}_{e}$
to be the same regardless of optical activity. However, the sensitivity
of splitting to input polarization has been modified dramatically.
Without gyrotropy there is a clear symmetric picture with peak maxima
corresponding to the cases when the incident wave coincides with one
of the eigenpolarizations (figure~\ref{fig:gyro_split}a). When gyrotropy
is present, the peak maxima become shifted by about 30~degrees, and
the transmittance at maximum does not reach unity (figure~\ref{fig:gyro_split}b).

To proceed with in-depth analysis, let us look at the transmitted
wave polarization (figure~\ref{fig:gyro_output}). The diagrams are
plotted for the same incident wave polarizations as in figure~\ref{fig:biiso_output}
at the frequencies of polarization-split doublet. One can see immediately
that the transmitted wave polarization is modified considerably if
gyrotropy is present. The structure still works as a polarizer in
a sense that at both peaks the transmitted wave is always polarized
similarly for any input polarization. This can be seen by comparing
the graphs in figure~\ref{fig:gyro_output}a-c as well as in figure~\ref{fig:gyro_output}d-f.
However, the output polarization states are found to depend on the
optical activity strength present. This dependence mainly manifests
itself as rotation, so the output polarization states remain visibly
orthogonal for the two peaks. However, besides the change in orientation
there is also a change in ellipticity, as can best be seen in figure~\ref{fig:gyro_output}f.

\begin{figure}
\includegraphics[%
  width=0.50\linewidth]{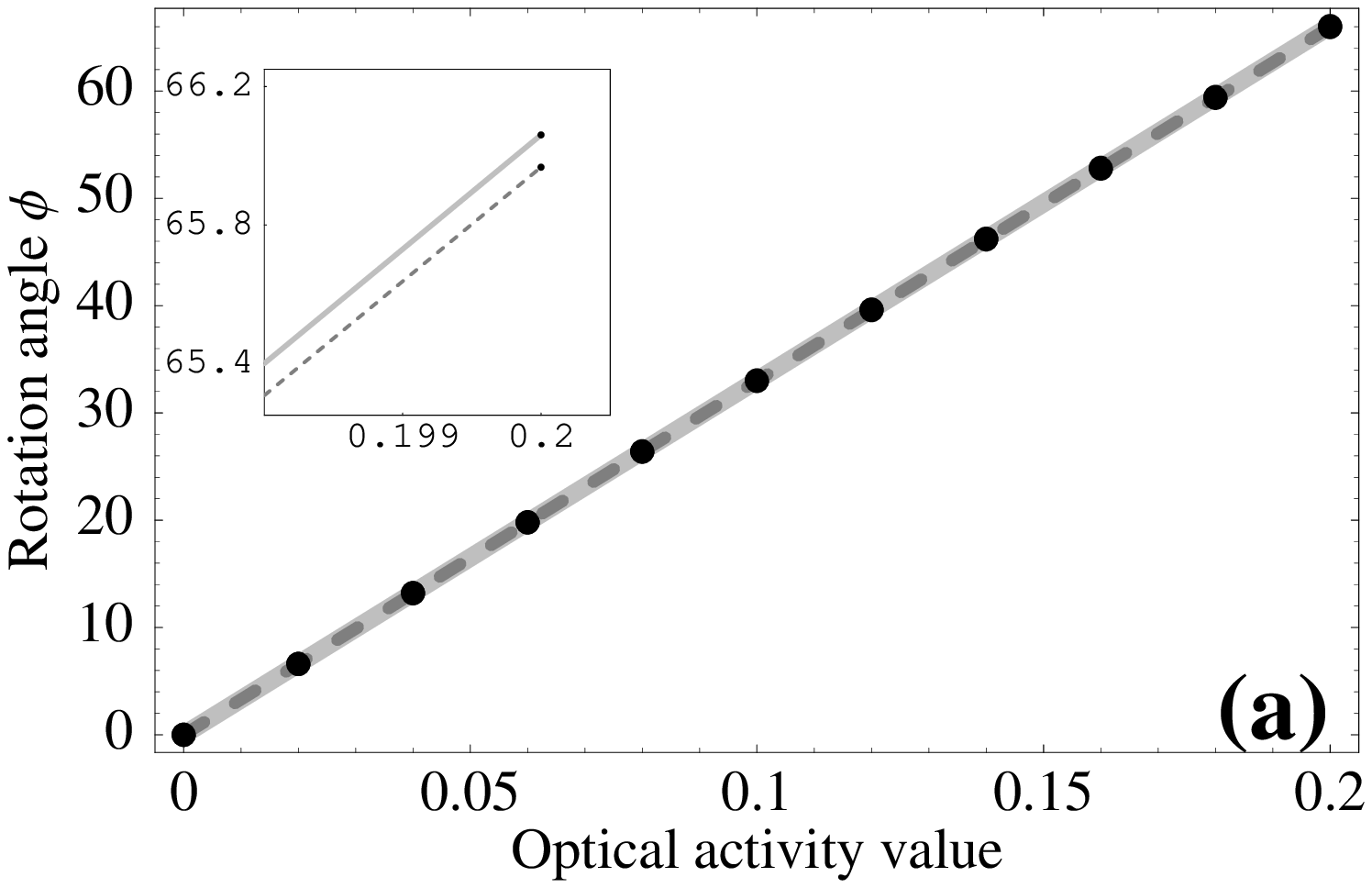}\includegraphics[%
  width=0.50\linewidth]{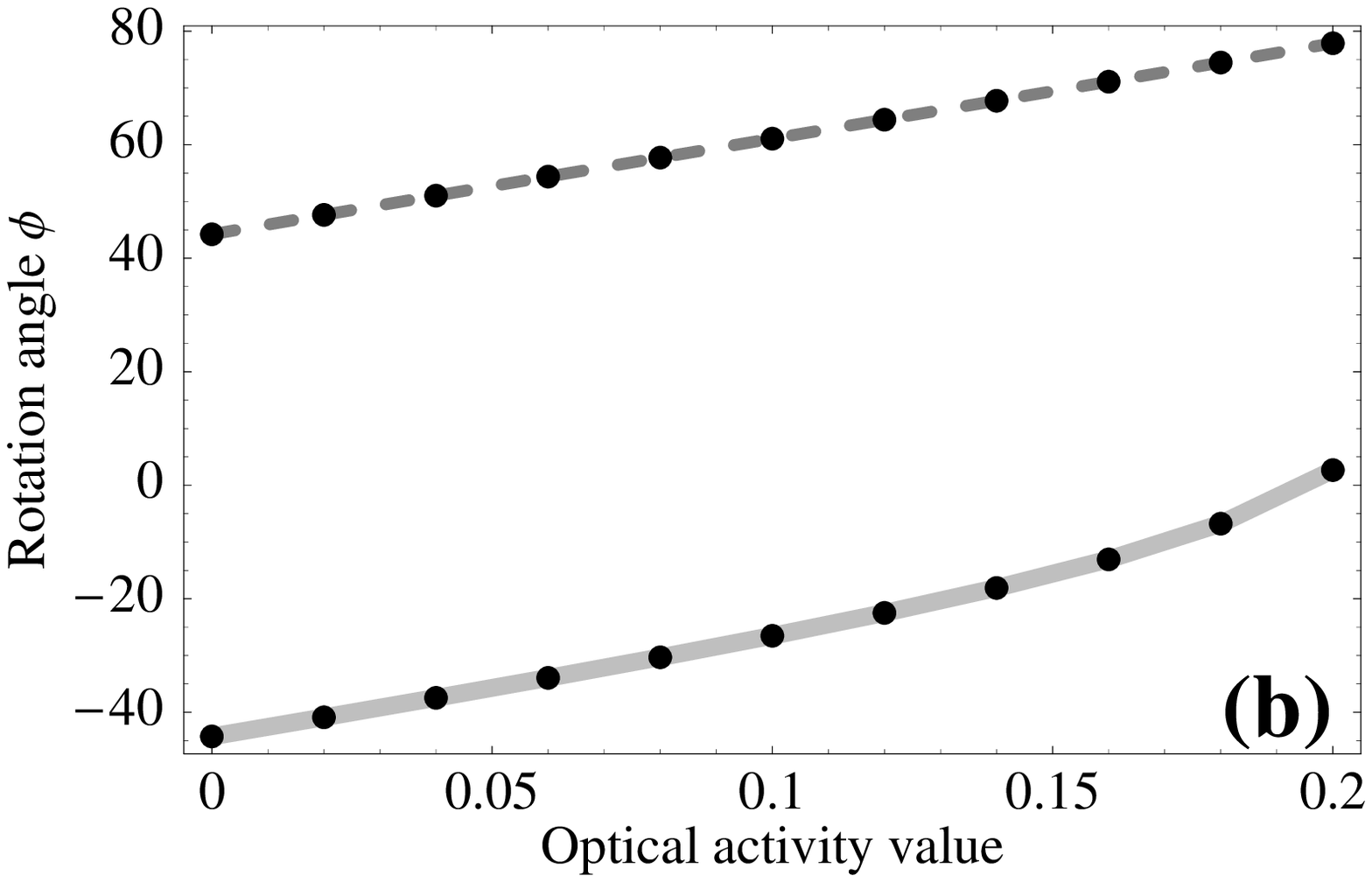}

\caption{\label{fig:rotation_comp}The dependence of polarization rotation
angle~$\phi$ on the value of~$\alpha$ for isotropic \textbf{(a)}
and birefringent \textbf{(b)} multilayer structure. The incident wave
polarization is according to Fig.~\ref{fig:gyro_output}c for the
split peak frequencies $\omega/\omega_{0}=1.0923$ (\emph{dashed})
and $\omega/\omega_{0}=1.0938$ (\emph{solid}). The inset shows a
slight difference in the slope for the two lines according to~(\ref{eq:rotation}).}
\end{figure}

It is also worth noting that the character of polarization rotation
in this case is completely different from what was observed in the
previous section. This difference is best seen comparing figure~\ref{fig:biiso_output}c,f
and figure~\ref{fig:gyro_output}c,f, when the same incident wave
polarization is considered, and plotting the angle of rotation~$\phi$
versus optical activity~$\alpha$ (figure~\ref{fig:rotation_comp}).
In the bi-isotropic case the rotation is always with respect to \emph{incident
wave}, it is independent of the structure's geometrical or spectral
features, so~$\phi$ is almost equal for resonant and off-resonant
frequencies, exhibiting only a slight frequency variation according
to~(\ref{eq:rotation}) (figure~\ref{fig:rotation_comp}a).

In the case when isotropic optical activity is added to birefringence,
the dominant feature is polarization-induced splitting, its polarization
properties subject to modification due to gyrotropy. This happens
in a multilayer because splitting (along with all spectral features)
is resonant and the more layers there are in a multilayer, the more
pronounced the peaks are. On the other hand, gyrotropy-induced rotation
in~(\ref{eq:p_gyro_1}) only weakly increases with the number of
layers. So, the rotation occurs with respect to \emph{eigenpolarizations}
of the birefringent medium. The angle of rotation~$\phi$ is thus
strongly dependent on frequency, exhibiting a change of $90^{\circ}$
over a very small change ($\Delta\omega/\omega_{0}=0.0015$) between
two polarization-split peaks (figure~\ref{fig:rotation_comp}b).
The dependence $\phi(\alpha)$ itself no longer conforms to~(\ref{eq:rotation})
and even exhibits a slight nonlinearity, apparently connected to the
emerging ellipticity in output polarization (see figure~\ref{fig:gyro_output}f).

It may appear at first that this change of output polarization should
correspond to the change of polarization eigenstates in gyrotropic
media. However, in this case one would expect exactly the same propagator
structure as in~(\ref{eq:p_uniaxial}) with~$\mathbf{a}_{0}$~and~$\mathbf{b}_{0}$
replaced with elliptical eigenvectors, which could have been recoverable
from figure~\ref{fig:gyro_output}. This, however, is not straightforward
(see~\ref{sec:app_a}), and this would also contradict with the structure
in~(\ref{eq:p_gyro_2}) where a term proportional to~$d\alpha$
is present. Together with an additional term in~(\ref{eq:p_gyro_1})
this means that eigenvectors should be thickness-dependent. This cannot
happen since eigenvectors of~$\mathbb{P}$ are the same as those
of~$\mathbb{M}$, which depends only on material parameters.

Indeed, if only change of eigenpolarizations were involved here, then
gyrotropy would leave unchanged the Malus-like dependence of transmittance
on input polarization, with respect to rotated output polarization
states -- quite contrary to what is seen in figure~\ref{fig:gyro_output}c
where the output and input waves are almost mutually perpendicular
in polarization, yet the transmittance is near unity. So one may guess
that there should be an interplay between ``bi-isotropic-like'' uniform
polarization rotation and gyrotropy-induced change of eigenpolarizations.
The exact nature of this interplay is yet to be understood.

To summarize this section, we have found out that isotropic optical
activity added to a birefringent multilayer structure influences both
the transmission spectra and the transmitted wave polarization. However,
the modification of the spectra can be seen in the change of polarization
sensitivity. The nature of polarization-induced splitting, the frequencies
of split components, and the orthogonality of output polarization
for these components remains unaltered.

The results obtained can be used to control the output polarization
of a splitter based on an NPD multilayer with anisotropic layers.
This control occurs without any modification to the splitter frequency
characteristics. This is definitely of interest in the design of integrated
optical polarizers, as both polarization directions can be rotated
without having to physically re-orient the multilayer. Again, fractal
structures are geometrically preferable as NPD structures because
peak multiplicity allows polychromatic filtering properties. This
can facilitate both multiple-channel polarization separators and split
component matching, which can in turn be promising in applications.

\section{Conclusions and outlook\label{sec:Discussion}}

In conclusion, we have analyzed the influence of isotropic optical
activity, the simplest kind of gyrotropy, on the spectral and polarization
properties of deterministic non-periodic (in particular, fractal)
multilayer structures. In \emph{isotropic} case, we have shown numerically
and analytically that optical activity does not cause any change in
the transmission spectrum, but rather leads to a uniform polarization
rotation. It is independent of the structure geometry and can be described
by~(\ref{eq:rotation}). In \emph{birefringent} case, we have shown
that the frequencies of polarization-split doublet do not change.
However, both the output polarization and the value of transmittance
at peak frequencies do depend on the strength of gyrotropy. In more
complicated cases, when gyrotropy-related terms are expected to cross-couple
with $\varepsilon$~and~$\mu$ in the single-layer characteristic
operator~(\ref{eq:abcd_n}--\ref{eq:v1234_n}), more significant
modification of the spectra can be expected, e.g., changes of resonant
peak locations. 

The results obtained are characteristic to any structure geometry
that facilitates sharp resonance peaks located in a band gap, as is
the case with many deterministic non-periodic media. Fractal geometry
considered here allows the effects to be manifest in the desire frequency
range with desired peak multiplicity. Among other notable geometries
one can also name single- and multiple-cavity Fabry-P\'erot resonators
embedded into a periodic multilayer.

The effects observed can be used in the design of frequency-selective,
compact devices for polarization-sensitive integrated optics. This
particularly concerns mono- and polychromatic absorptionless polarizers
with controlled output polarization. Devices that rotate the polarization
of output beam with respect to the input are also possible. Note,
however, that in this paper no account was taken for present-day experimental
or technological applicability of the results obtained. Also note
that the combination of parameters described in Section~\ref{sec:Gyrotropy}
is not among those naturally occurring in crystals (though it may
be possible in nanocomposites). Nevertheless, this combination is
very useful illustratively, since it allows analytical treatment and
so can be used as a landmark of what will happen if we complicate
the system further.

There is a broad range of problems to be addressed within the scope
of extending the present research. First and foremost, it appears
fruitful to investigate more complicated cases of optical anisotropy
and gyrotropy. This is especially relevant is two cases. First, of
interest are materials with properties subject to external tuning,
e.g., liquid crystals or Pockels, Faraday, or Kerr media. Secondly,
as was mentioned above, of importance are anisotropic and/or gyrotropic
materials which can be readily used in multilayer micro- and nanostructure
fabrication. Since the numerical techniques used in this paper are
based on the general material equations~(\ref{eq:mat}), it appears
possible to rigorously allow for all kinds of anisotropy and gyrotropy,
as well as to account for polarization effects during wave propagation.
This paper demonstrates that analytical or semi-analytical treatment
is possible using symbolic computation techniques. A thorough classification
of spectral and polarization-related effects resulting from anisotropy
and/or gyrotropy according to different crystallographic symmetries
existent in nature is worthwhile, too. 

Other areas of extending this paper are also apparent. Namely, it
is interesting to find out what happens in NPD multilayers containing
anisotropic or gyrotropic layers of the same material but different
spatial orientation. In addition, it also appears promising to combine
materials with various optical properties in different parts of the
same NPD multilayer. Since such structures generally tend to exhibit
eigenmodes with distinct localization patterns, one may expect that
by introducing different materials into different localization regions
one can produce devices with unusual and largely tunable spectral
dependencies of optical properties.

\begin{ack}
The authors wish to acknowledge helpful discussions with S.~Kurilkina,
A.~Smirnov, G.~Borzdov, L.~Burov, A.~Novitsky, and A.~Lavrinenko.
This work was supported in part by the Basic Research Foundation of
Belarus (Grant No.~F04M--140) as well as by Deutsche Furschungsgemeinschaft
(DFG SPP 1113).
\end{ack}
\appendix

\section{Bi-isotropic media and circular polarization\label{sec:app_a}}

From the reasoning of Section~\ref{sec:Biisotropic}, one may conceive
that bi-isotropic media allegedly do not discriminate between left-handed
and right-handed circular polarization, while it is commonly known
(see \cite{Georgieva,BarkovskJMO}) that on the contrary, change of
effective refractive index occurs. Here we would like to explain in
detail how and why this effective index change does \emph{not} cause
any polarization splitting.

We take a single bi-isotropic layer and analyze what happens to a
circularly polarized incident wave. Starting with~(\ref{eq:w_biiso}),
we can see that in this case ($\phi\equiv kd\alpha$)\[
\mathbf{H}_{0}^{+}=H_{0}(\mathbf{a}_{0}+i\mathbf{b}_{0}),\]
\begin{eqnarray}
\mathbb{W}^{+} & = & H_{0}\mathbb{L}\left[\begin{array}{c}
1\\
\frac{1}{n_{0}}\end{array}\right]\left[\left(\cos\phi+i\sin\phi\right)\mathbf{a}_{0}+i\left(\cos\phi+i\sin\phi\right)\mathbf{b}_{0}\right]=\label{eq:ap_w_circ1}\\
 & = & \mathbb{L}\left[\begin{array}{c}
1\\
\frac{1}{n_{0}}\end{array}\right]\mathbf{H}_{0}^{+}e^{i\phi}.\nonumber \end{eqnarray}

We can see that there is no change of polarization but rather a change
of phase for the transmitted wave with respect to that for isotropic
layer. The value of this change is~$\phi$. If one reverses the direction
of circular polarization, one gets\[
\mathbf{H}_{0}^{-}=H_{0}(\mathbf{a}_{0}-i\mathbf{b}_{0}),\]
\begin{eqnarray}
\mathbb{W}^{-} & = & H_{0}\mathbb{L}\left[\begin{array}{c}
1\\
\frac{1}{n_{0}}\end{array}\right]\left[\left(\cos\phi-i\sin\phi\right)\mathbf{a}_{0}+i\left(-\cos\phi+i\sin\phi\right)\mathbf{b}_{0}\right]=\label{eq:ap_w_circ2}\\
 & = & \mathbb{L}\left[\begin{array}{c}
1\\
\frac{1}{n_{0}}\end{array}\right]\mathbf{H}_{0}^{-}e^{-i\phi}.\nonumber \end{eqnarray}

As can be noticed, the change of phase is different in this case.
A conventional approach with bulk bi-isotropic media is to combine
this phase shift~$\phi$ with propagation phase $\varphi=kd\sqrt{\epsilon\mu}$.
This results in a change in effective refractive index, such that
\cite{Georgieva}\begin{equation}
n_{\textnormal{eff}}^{\pm}=n_{0}\pm\delta n=\sqrt{\epsilon\mu}\pm\alpha.\label{eq:ap_n_eff}\end{equation}

However, our studies show that such an approach can yield misleading
results when applied even to a single layer. A straightforward conclusion
of the effective index concept would be that the layer, and even more
so the multilayer structure, should become sensitive to the orientation
of circular polarization. Indeed, as spectral features strongly depend
on~$n$, the difference in~$n_{\textnormal{eff}}$ should result
in the difference in the spectra, causing polarization-induced splitting
with respect to circular eigenpolarizations. 

However, this contradicts with both numerical and analytical results.
Indeed, note first that the phase change in~(\ref{eq:ap_w_circ1}--\ref{eq:ap_w_circ2}),
even if included in~$\mathbb{L}$, does not contribute to transmittance.
Further, considering a multilayer composed of~$N$ isotropic and
bi-isotropic layers, one can use~(\ref{eq:mult_ii}--\ref{eq:mult_bb})
to see that\begin{equation}
\mathbb{W}^{\pm}=\mathbb{L}_{1}\mathbb{L}_{2}\cdots\mathbb{L}_{N-1}\mathbb{L}_{N}\left[\begin{array}{c}
1\\
\frac{1}{n_{0}}\end{array}\right]\mathbf{H}_{0}\exp(\pm\phi_{1}\pm\phi_{2}\pm\ldots\pm\phi_{N}).\label{eq:ap_w_arbitrary}\end{equation}

Since transmission spectrum is determined by the matrix product in~(\ref{eq:ap_w_arbitrary}),
which is polarization-independent, bi-isotropy in any of the constituent
layers cannot lead to polarization-induced splitting. Instead, it
contributes to an overall uniform phase change -- similar to a uniform
polarization rotation in the case of linear input polarization.

This example is a good confirmation of the fact that the refractive
index itself is no longer straightforward nor safe to use once complex
optical media are involved. Indeed, an effective refractive index
for a certain eigenpolarization essentially means that a wave with
this polarization propagates as if the medium were isotropic. But
it can only be true if the wave polarization is never changed during
propagation. In \emph{homogeneous} media of any kind, it is always
true by definition of an eigenwave \cite{BarkovskJMO} -- so it is
enough to make sure that the \emph{incident} wave is properly polarized,
which is what is conventionally assumed. 

However, in \emph{inhomogeneous} media this is no longer enough. Even
in a multilayer the layer interfaces can and do introduce polarization
coupling between eigenwaves. And this is exactly what occurs in a
bi-isotropic layer -- both layer interfaces serve as perfect couplers
between eigenpolarizations, reversing the polarization after each
reflection. (This cross-coupling is mentioned in \cite{JaggardSun}
\textbf{}and is due to the fact that right- or left-handedness of
a circularly-polarized wave is determined by the sign of the vector
product $i\mathbf{n}[\mathbf{H}\times\mathbf{H}^{*}]$ \cite{FedorovEng},
and during Fresnel reflection in our case $\mathbf{n}\rightarrow-\mathbf{n}$,~$\mathbf{H}\rightarrow r\mathbf{H}$,
$r$~being real). As such, even in the case of one layer, even in
the case of normal incidence and axial symmetry of the system, the
wave \emph{initially} polarized along one of the eigenwaves \emph{does
not retain} its polarization during propagation. Therefore the resulting
spectral properties even of one layer differs greatly from what would
have happened if one blindly used the effective index concept suitable
for homogeneous media.

\section{Evolution operator for a uniaxial, isotropically optically active
layer\label{sec:app_b}}

Here we would like to show in detail the way we have derived the model
propagator for a uniaxially birefringent layer with isotropic gyrotropy
to be used in~(\ref{eq:p_gyro_1}) and~(\ref{eq:p_gyro_2}). 

The main difficulty in evaluating the propagator directly from~(\ref{eq:m_gyro})
consists in the fact that eigenvectors of such a matrix are complex
elliptical vectors whose orientation and ellipticity depend on numerical
relations between~$\epsilon_{o}$,~$\epsilon_{e}$, and~$\alpha$,
and layer interfaces even for one layer will play a role not clearly
understood (see~\ref{sec:app_a}). The expression for~$\mathbb{P}$
becomes excessively complex to be analyzed successfully.

One of the ways out would be to construct approximations assuming
that~$\alpha$ and $\left|\epsilon_{e}-\epsilon_{o}\right|$ are
small quantities. However, such approximation has to be done carefully,
since while these quantities are small enough indeed, they appear
in the propagator multiplied by~$d\omega/c$, and the resulting phases
are no longer infinitesimal. So it can be misleading to straightforwardly
suppose $\left|\varepsilon_{e}-\varepsilon_{o}\right|^{2}=0=\alpha^{2}$.

Instead, what can be done is to analyze several orders of Taylor series
for~$\mathbb{P}$ in order to recover how the Taylor decomposition
can fold back into trigonometric expressions \emph{so as to produce
the same Taylor series up to a certain order}. The result of this
operation is what we call here the \emph{model propagator}. Reduced
to a $4\times4$ form {[}see the footnote for~(\ref{eq:p_biiso}){]},
it reads\begin{equation}
\mathbb{P}^{(g)}=\left[\begin{array}{cc}
A & B\\
C & D\end{array}\right],\label{eq:ap_p_gyro_general}\end{equation}
\begin{equation}
A\approx\left[\begin{array}{cc}
\cos kd\sqrt{\mu\epsilon_{o}}\cos\phi & \cos kd\sqrt{\mu\epsilon'}\left(1-\Delta\right)\sin\phi\\
-\cos kd\sqrt{\mu\epsilon'}\left(1+\Delta\right)\sin\phi & \cos kd\sqrt{\mu\epsilon_{e}}\cos\phi\end{array}\right],\label{eq:ap_p_gyro_a}\end{equation}
\begin{equation}
D\approx\left[\begin{array}{cc}
\cos kd\sqrt{\mu\epsilon_{o}}\cos\phi & \cos kd\sqrt{\mu\epsilon'}\left(1+\Delta\right)\sin\phi\\
-\cos kd\sqrt{\mu\epsilon'}\left(1-\Delta\right)\sin\phi & \cos kd\sqrt{\mu\epsilon_{e}}\cos\phi\end{array}\right],\label{eq:ap_p_gyro_d}\end{equation}
\begin{equation}
B\approx\left[\begin{array}{cc}
\frac{i\epsilon}{\sqrt{\mu\epsilon_{e}}}\sin kd\sqrt{\mu\epsilon_{o}}\cos\phi & \frac{i\epsilon}{\sqrt{\epsilon\mu}}\sin kd\sqrt{\epsilon\mu}\sin\phi\\
-\frac{i\epsilon}{\sqrt{\epsilon\mu}}\sin kd\sqrt{\epsilon\mu}\sin\phi & \frac{i\epsilon}{\sqrt{\mu\epsilon_{o}}}\sin kd\sqrt{\mu\epsilon_{e}}\cos\phi\end{array}\right],\label{eq:ap_p_gyro_b}\end{equation}
\begin{equation}
C\approx\left[\begin{array}{cc}
\frac{i\epsilon}{\sqrt{\mu\epsilon_{o}}}\sin kd\sqrt{\mu\epsilon_{o}}\cos\phi & \frac{i\mu}{\sqrt{\epsilon\mu}}\sin kd\sqrt{\epsilon\mu}\sin\phi\\
-\frac{i\mu}{\sqrt{\epsilon\mu}}\sin kd\sqrt{\epsilon\mu}\sin\phi & \frac{i\mu}{\sqrt{\mu\epsilon_{e}}}\sin kd\sqrt{\mu\epsilon_{e}}\cos\phi\end{array}\right],\label{eq:ap_p_gyro_c}\end{equation}

In order to simplify off-diagonal terms to the maximum possible extent,
we have substituted $\epsilon_{o}\equiv\epsilon-\delta\epsilon$,
$\epsilon_{e}\equiv\epsilon+\delta\epsilon$. Also, $\Delta\equiv\frac{1}{2}\sin(\delta\epsilon/\epsilon)$
and $\epsilon'\equiv\epsilon\left(1-\delta\varphi\right)$ where $\delta\varphi\equiv\delta\epsilon(kd\sqrt{\epsilon\mu})^{-2}$,
and as before, $\phi=kd\alpha$.

It can be confirmed that the propagator constructed according to~(\ref{eq:ap_p_gyro_general}--\ref{eq:ap_p_gyro_c})
yields the same Taylor decomposition as the real matrix exponential
when the terms up to~$\alpha^{2}$ and~$\delta\epsilon^{2}$ are
preserved. 

Certainly, the propagator of~(\ref{eq:ap_p_gyro_general}--\ref{eq:ap_p_gyro_c})
is by no means a true form of the evolution operator; it is not even
an approximation in the strict sense -- rather, it is an expression,
which yields the same approximation. Hence it can be used as an approximate
model of a real gyrotropic layer. The fact that it is trigonometrical
allows us to analyze such phenomena as polarization plane rotation
and birefringence, reverting to decomposition only when and where
needed. The only apparent limitation of this approach is that we should
not decompose~(\ref{eq:ap_p_gyro_a}--\ref{eq:ap_p_gyro_c}) to more
than the second order with respect to~$\alpha$ and~$\delta\epsilon$.

This kept in mind, one can take Taylor in square roots with respect
to~$\delta\epsilon$ and neglect the product~$\alpha\delta\epsilon$
(which is essentially a quadratic term). This yields ($\varphi\equiv kd\sqrt{\epsilon\mu}$)\begin{equation}
\begin{array}{l}
A=D\approx P\cos\varphi+\delta\epsilon\lambda_{d}S\sin\varphi,\\
B\approx\frac{i\epsilon}{\sqrt{\varepsilon\mu}}\left[P\sin\varphi-\delta\epsilon\left(\lambda_{od}+\lambda'\right)S\cos\varphi\right],\\
C\approx\frac{i\mu}{\sqrt{\epsilon\mu}}\left[P\sin\varphi+\delta\epsilon\left(\lambda_{od}-\lambda'\right)S\cos\varphi\right],\\
P\equiv I\cos\phi-\mathbf{q}^{\times}\sin\phi,\quad S\equiv\mathbf{a}_{0}\otimes\mathbf{a}_{0}-\mathbf{b}_{0}\otimes\mathbf{b}_{0}\end{array}\label{eq:ap_abcd_decomp}\end{equation}
where the subscripts~$d$~and~$od$ stand for {}``diagonal''
and {}``off-diagonal'', and \begin{equation}
\lambda_{d}=\frac{kd\mu}{2\sqrt{\epsilon\mu}},\quad\lambda_{od}=\frac{kd\sqrt{\epsilon\mu}}{2\epsilon},\quad\lambda'=\frac{1}{2\epsilon}.\label{eq:ap_abc_lambdas}\end{equation}

Therefore one can rewrite~(\ref{eq:ap_abcd_decomp}--\ref{eq:ap_abc_lambdas})
in the form analogous to~(\ref{eq:p_short}) and~(\ref{eq:p_uniaxial}):\begin{equation}
\mathbb{P}^{(g)}\approx\mathbb{L}\left(I\cos\phi-\mathbf{q}^{\times}\sin\phi\right)+\delta\epsilon\mathbb{L}'\left(\mathbf{a}_{0}\otimes\mathbf{a}_{0}-\mathbf{b}_{0}\otimes\mathbf{b}_{0}\right)\label{eq:ap_p_gyro_structure_1}\end{equation}
Eq.~(\ref{eq:ap_p_gyro_structure_1}) is used in the text as~(\ref{eq:p_gyro_1}).
The coefficient matrices are

\begin{equation}
\mathbb{L}=\left[\begin{array}{cc}
\cos\varphi & \frac{i\epsilon}{\sqrt{\epsilon\mu}}\sin\varphi\\
\frac{i\mu}{\sqrt{\epsilon\mu}}\sin\varphi & \cos\varphi\end{array}\right],\label{eq:ap_gyro_matrix_1}\end{equation}

\begin{equation}
\mathbb{L}'=\left[\begin{array}{cc}
\frac{kd\mu}{2\sqrt{\epsilon\mu}}\sin\varphi & -\frac{i\epsilon}{\sqrt{\epsilon\mu}}\left(\frac{kd\sqrt{\epsilon\mu}}{2\epsilon}+\frac{1}{2\epsilon}\right)\cos\varphi\\
\frac{i\mu}{\sqrt{\epsilon\mu}}\left(\frac{kd\sqrt{\epsilon\mu}}{2\epsilon}-\frac{1}{2\epsilon}\right)\cos\varphi & \frac{kd\mu}{2\sqrt{\epsilon\mu}}\sin\varphi\end{array}\right].\label{eq:ap_gyro_matrix_2}\end{equation}

On the other hand, one can take Taylor of~(\ref{eq:ap_p_gyro_a}--\ref{eq:ap_p_gyro_c})
in another manner, excluding the propagation phase~$\varphi$ rather
than the polarization plane rotation~$\phi$ from decomposition.
Making use of the fact that \[
\frac{i\epsilon}{\sqrt{\mu\left(\epsilon\mp\delta\epsilon\right)}}\sin kd\sqrt{\mu\left(\epsilon\pm\delta\epsilon\right)}\simeq\frac{i\left(\epsilon\pm\delta\epsilon\right)}{\sqrt{\mu\left(\epsilon\pm\delta\epsilon\right)}}\sin kd\sqrt{\mu\left(\epsilon\pm\delta\epsilon\right)}+O\left(\delta\epsilon^{2}\right)\]
and likewise neglecting~$\alpha\delta\epsilon$, we obtain

\begin{equation}
\begin{array}{l}
A=D\approx\mathbf{a}_{0}\otimes\mathbf{a}_{0}\cos\varphi_{o}+\mathbf{b}_{0}\otimes\mathbf{b}_{0}\cos\varphi_{e}-\alpha kd\cos\varphi\mathbf{q}^{\times},\\
B\approx\frac{i\epsilon_{o}}{\sqrt{\epsilon_{o}\mu}}\mathbf{a}_{0}\otimes\mathbf{a}_{0}\sin\varphi_{o}+\frac{i\epsilon_{e}}{\sqrt{\epsilon_{e}\mu}}\mathbf{b}_{0}\otimes\mathbf{b}_{0}\sin\varphi_{e}-\frac{i\epsilon}{\sqrt{\varepsilon\mu}}\alpha kd\sin\varphi\mathbf{q}^{\times},\\
C\approx\frac{i\mu}{\sqrt{\varepsilon_{o}\mu}}\mathbf{a}_{0}\otimes\mathbf{a}_{0}\sin\varphi_{o}+\frac{i\mu}{\sqrt{\varepsilon_{e}\mu}}\mathbf{b}_{0}\otimes\mathbf{b}_{0}\sin\varphi_{e}-\frac{i\mu}{\sqrt{\varepsilon\mu}}\alpha kd\sin\varphi\mathbf{q}^{\times}.\end{array}\label{eq:ap_abcd_decomp_uniax}\end{equation}

Here, naturally, $\varphi_{o}\equiv kd\sqrt{\varepsilon_{o}\mu}$,
$\varphi_{e}\equiv kd\sqrt{\varepsilon_{e}\mu}$. Thus, isolating
the real-space structure on analogy with~(\ref{eq:ap_p_gyro_structure_1}),
we have

\begin{equation}
\mathbb{P}^{(g)}\approx\mathbb{L}_{o}\mathbf{a}_{0}\otimes\mathbf{a}_{0}+\mathbb{L}_{e}\mathbf{b}_{0}\otimes\mathbf{b}_{0}-kd\alpha\mathbb{L}\mathbf{q}^{\times}.\label{eq:ap_p_gyro_structure_2}\end{equation}

This formula is used in the text as~(\ref{eq:p_gyro_2}). The coefficient
matrices~$\mathbb{L}_{a}$ and~$\mathbb{L}_{b}$ are identical to
those in~(\ref{eq:lambda_uniax}). The antisymmetric term contains
the same matrix~$\mathbb{L}$ as in~(\ref{eq:ap_gyro_matrix_1}).

\end{document}